\documentclass{article}

\usepackage[dblblindworkshop,final]{neurips_2026}
\workshoptitle{TAE (Trust-AI-Eval): Can We Trust AI Evaluation?}

\usepackage[T1]{fontenc}
\usepackage[utf8]{inputenc}
\usepackage{microtype}
\usepackage{inconsolata}
\usepackage{graphicx}
\usepackage{subcaption}
\usepackage{CJKutf8}
\usepackage{amsmath,amssymb}
\usepackage{dsfont}  %
\usepackage{float}
\usepackage{listings}
\usepackage{xcolor}
\usepackage{xspace}
\usepackage{multirow}
\usepackage{arydshln}
\usepackage{url}
\usepackage[hidelinks]{hyperref}

\newcommand{\cjk}[1]{\begin{CJK}{UTF8}{gbsn}#1\end{CJK}}
\title{PyMETA: Evaluating Student Code Diagnosis on and Beyond the First Execution Error}
\author{%
  Ziqi Tang\thanks{Equal contribution.} \quad
  Chuyue Li\footnotemark[1] \quad
  Jingyi Wang \quad
  Yu Wu \quad
  Kazuma Hashimoto \quad
  Lingyu Gao \\
  CircleCat \\
  \texttt{\{cyli, ztang, jasmine, wuyu, kazumah, lygao\}@circlecat.org}
}

\newcommand{\datasetname}{\textsc{PyMETA}\xspace}

\begin{document}
\maketitle

\begin{abstract}
Code-diagnosis evaluation depends on how error supervision is constructed and
what that supervision treats as a correct diagnosis. We introduce \datasetname,
a dataset of 48,646 Python submissions with execution-derived single-error labels, and a targeted subset of 97 submissions annotated by experts through repair and re-execution. Under the first-outcome protocol, four recent prompted models obtain 87.5--93.8\% macro F1, all above the strongest finetuned baseline at 80.6\%. This reverses the comparison obtained with the earlier set of prompted models. The expert subset gives a less favorable account of complete diagnosis: exact-set match ranges from 43.3\% to 48.5\%, although sample F1 is approximately 79--81\%. The evaluation setup also changes which behavior is visible. On the same 45 audited submissions without an expert-annotated \texttt{Logic Error}, none of the four recent models predicts that label when restricted to one output, whereas 46.7--57.8\% include it under multi-error prompting, usually after an explicit-error label. These findings show that model comparisons and error-bias claims depend on what the gold label represents, which outputs the model is allowed to return, and how those outputs are scored.
\end{abstract}

\section{Introduction}
Automated programming assessment often derives feedback from running tests, and
that feedback is frequently limited to whether tests pass or fail and how the
output differs from a reference solution~\citep{Messer2024}. In this setting, one
reproducible target for student-code diagnosis is the first outcome an Online
Judge exposes. This target is inexpensive at scale and aligned with
execution-system feedback, but it also defines what counts as a correct diagnosis.
A model that reads the whole submission may instead identify an error that becomes
observable only after the current failure is repaired, and under first-outcome
scoring such a prediction is penalized in the same way as one that is simply
wrong.

The distinction is consequential even within compiler feedback: analyzing over 21
million novice compiler messages, \citet{Becker2018First} find that the ranks and
frequencies of error types differ when only first messages are counted. The
subsequent diagnostics studied by Becker et al. are emitted within the same
compilation; in our setting some error categories are not observable in the
current execution at all, emerging only after repair and re-execution. The
boundary matters as models identify problems beyond recorded judge outcomes.
Under test-time scaling, the best-performing models in UOJ-Bench uncover errors
in over 5\% of submissions an Online Judge had scored as fully
correct~\citep{uojbench2026}. For categorical diagnosis of student programs the question becomes:
when a prediction differs from the first outcome, is it unsupported, or does it
name an error exposed later along a repair path?

We introduce \datasetname\ (Python Multi-Error TAxonomy), a benchmark built to make that distinction measurable:
48,646 real Python submissions from 153 problems under a hierarchical error
taxonomy, each labeled by the first outcome exposed by Online Judge execution. The submissions come from Circle Cat, a 501(c)(3) non-profit that provides free software-engineering education to Chinese-speaking women; 579 learners wrote them during ordinary coursework on the organization’s Moodle-based Online Judge. 
For a targeted subset of 97 submissions, selected from CodeBERT misclassifications
and high-entropy predictions, experts repeatedly observed an error, repaired it,
and re-executed the program, producing repair-conditional error sets. Pairing
these with the first-outcome labels lets the same prediction be evaluated against
two related but distinct targets. These sets are not assumed exhaustive, and the
subset is targeted rather than random, so it characterizes how the two targets
can diverge rather than how often they do so.

Using \datasetname, we study three research questions. \textbf{RQ1} asks how models perform
under the standard first-outcome protocol. On the same test set, with the
single-error prompt unchanged across the prompted models, the best finetuned
baseline, CodeLlama-7B, reaches 80.6\% macro F1 while four recent prompted models
reach 87.5--93.8\%, reversing the model-selection conclusion supported by the
earlier set of prompted models. \textbf{RQ2} asks whether predictions that disagree with the
first-outcome label are supported along a repair path. Coverage against the expert
sets is 50\% and 52\% in the two samples; in the misclassification-based sample,
where every prediction disagrees with the recorded label, half of those
disagreements are supported and half are not. \textbf{RQ3} asks how the evaluation
configuration changes what is visible. No recent model exceeds 50\% exact-set
match, and on 45 audited submissions with no expert-annotated \texttt{Logic
Error}, none of the four models predicts that label under single-error prompting,
whereas 46.7--57.8\% of their multi-error predictions include it.

Together, these results separate three notions that code-diagnosis evaluation can
conflate: reproducing the first outcome, naming an error supported somewhere along
a repair path, and recovering the complete repair-conditional set. \datasetname
supports all three, and the conclusions drawn depend on which target, output
space, and scoring rule define correctness.

\section{Related Work}
\subsection{Code Error Datasets and Label Construction}
Code error detection, classification, localization, and repair have long been
studied in programming education~\citep{11202742}; resources differ in how their
error supervision is constructed. \citet{Shirafuji_2023} infer multiple errors
from rule-based comparison of wrong and correct programs, capturing differences
between repair endpoints rather than failures exposed during execution.
COJ2022~\citep{10.1145/3539618.3591680} annotates semantic error types and
locations alongside repaired versions, while Online Judge datasets such as
ACcoding~\citep{accoding2024} preserve submission histories and judge outcomes
without fine-grained per-submission error sets. Each construction encodes a
different view of program errors, and an evaluation built on those labels judges
predictions against that particular view.

Model-identified problems can extend beyond recorded judge outcomes. UOJ-Bench
evaluates LLMs on code generation, hacking, and repair over real Online Judge
submissions~\citep{uojbench2026}; under test-time scaling, its best-performing
models uncover errors in over 5\% of submissions the original judge had scored as
fully correct. This raises a question for categorical diagnosis that no single
view can answer: when a prediction differs from the first observed outcome, is it
wrong, or does it identify an error exposed only after the current failure is
addressed? \datasetname makes this distinction measurable by pairing first-outcome
labels with repair-conditional annotations obtained through iterative repair and
re-execution on a targeted subset. These annotations are not assumed exhaustive,
but allow the same prediction to be evaluated against both targets.

\subsection{Model-Based Code Diagnosis}
Prior work improves code error classification through model architecture and training, including CodeT5 with ML-KNN~\citep{10602509}, finetuned BERT models for multi-label classification~\citep{10820190}, and BiLSTM--TextCNN models~\citep{10935456}. Other work studies logical-error diagnosis~\citep{lee2024improvingllmclassificationlogical} and the use of execution traces or execution feedback for program reasoning and code synthesis~\citep{wang2025semantics,gehring2025rlef}. Our focus is complementary. We compare finetuning and prompting but do not propose a new model architecture. Instead, we study how the gold-label definition, output cardinality, and scoring rule affect conclusions drawn from the same benchmark.

\section{PyMETA: Benchmark Design and Evaluation Setup}
\label{sec:dataset-construction}
This section defines the two evaluation targets used in the paper. The full benchmark measures agreement with the first outcome returned by the Online Judge, while the diagnostic subset records errors exposed through expert repair and re-execution. We describe the data, label taxonomy, and evaluation settings needed to keep these targets distinct.
\subsection{Dataset and Gold Labels}
\label{sec:dataset-overview}

The student submissions in \datasetname were collected from the online learning platform operated by Circle Cat Inc., a 501(c)(3) non-profit organization that provides free, accessible technical education and career-development guidance to Chinese-speaking women. Circle Cat's programs span three areas: structured software-engineering coursework that takes students from foundational concepts to advanced skills; a residency program in which students work on real-world open-source, non-profit, and industry projects under one-on-one mentorship; and career support from mentors with industry experience.

The coursework is delivered through a self-hosted Moodle instance with
an integrated Online Judge, where learners submit Python solutions
that are automatically compiled, executed, and graded with immediate
feedback. The historical submission logs from this platform form the
raw source of our dataset. Because
these submissions are produced by learners of varying proficiency
during ordinary coursework, they provide a realistic and diverse
distribution of student code errors.

\datasetname contains 48,646 Python submissions from 579 learners across 153 problems and 22 problem types. Each record includes the problem description, a reference solution, the student submission, and the compilation and execution results returned by Moodle CodeRunner. The data are split by \texttt{QuestionID}: 133/11/9 problems and 38,918/4,863/4,865 submissions in train/dev/test. A problem-level split prevents the same problem statement and its characteristic error patterns from appearing in both training and evaluation data. Dataset extraction and preprocessing are described in Appendix~\ref{sec:dataset-preprocess}.

The standard gold label is the first outcome exposed by the Online Judge. Parse-time failures are recorded before runtime failures; a runtime exception is recorded before any later behavior; and \texttt{Logic Error} is assigned when the program runs but fails one or more tests. This protocol is close to ordinary IDE feedback, but it does not claim to enumerate every fault in the submission.

The diagnostic subset contains 97 test submissions with expert multi-error labels. Annotators repeatedly executed each program, fixed one observed error at a time, and reran it until no explicit error remained; test failure after those repairs was labeled \texttt{Logic Error}. The resulting labels are therefore \emph{repair-conditional error sets}: some errors become observable only after an earlier error is repaired. The subset contains 187 labels in total (1.93 per submission) across 13 error types. Full annotation instructions appear in Appendix~\ref{sec:annotation-guidelines}.

The dataset and associated resources, including the splits, prompts, annotation guidelines, and evaluation scripts, are publicly available at \url{https://github.com/Circle-Cat/pymeta}; the dataset is also hosted on Hugging Face as \texttt{CircleCat/pymeta}.

\subsection{Error Taxonomy}
\label{sec:taxonomy-design}

The taxonomy has three nested levels. Task A distinguishes \texttt{No Error} from \texttt{Error}. Task B separates \texttt{No Error}, \texttt{Explicit Error}, and \texttt{Logic Error}. Task C uses 14 classes: \texttt{No Error}, \texttt{Logic Error}, 11 named Python exceptions, and \texttt{Other Errors}. The distinction between explicit and logic errors reflects the stage at which the Online Judge stops: explicit errors prevent successful completion, whereas logic errors are inferred from failed tests after execution.

The taxonomy is an evaluation instrument rather than a complete theory of student mistakes. It groups low-frequency exceptions and records system-observable outcomes, not pedagogical causes. Class definitions and the mapping from the raw annotation labels to the three levels are provided in Appendix~\ref{sec:appendix-taxonomy-defs} (Table~\ref{tab:taxonomy-mapping}); Table~\ref{tab:error-type-distribution} gives the full class counts.

\begin{table}[t]
  \centering
  \small
  \begin{tabular}{lr@{\qquad}lr}
    \hline
    \textbf{Class} & \textbf{Count} & \textbf{Class} & \textbf{Count} \\
    \hline
    No Error & 23,207 & Logic Error & 11,387 \\
    Syntax Error & 5,618 & Name Error & 2,565 \\
    Type Error & 2,074 & Indentation Error & 1,355 \\
    Unbound Local Error & 482 & Key Error & 417 \\
    Index Error & 400 & EOF Error & 277 \\
    Recursion Error & 334 & Value Error & 190 \\
    Tab Error & 162 & Other Errors & 178 \\
    \hline
  \end{tabular}
  \caption{Task C label distribution in the full \datasetname dataset.}
  \label{tab:error-type-distribution}
\end{table}

\subsection{Evaluation Setup}
\label{sec:single-error-finetuning}
We evaluate two related tasks. The single-error task asks for the first execution-grounded label, whereas the multi-error task asks for an unordered set of labels supported along the expert repair path. Because the tasks use different prompts and output spaces, results are interpreted within each task.

We evaluate two finetuned baselines, CodeBERT~\citep{feng2020codebertpretrainedmodelprogramming} and CodeLlama-7B~\citep{roziere2024codellamaopenfoundation}, together with two sets of prompted models. The earlier set includes GPT-3.5~\citep{openai2023gpt35turbo}, GPT-4o~\citep{openai2024gpt4osystemcard}, Gemini 2.5 Pro~\citep{sundar2023gemini}, and DeepSeek-V3~\citep{DeepSeekV3_2024}. The recent set includes GPT-5.5~\citep{openai2026gpt55}, Gemini 3.1 Pro Preview~\citep{google2026gemini31pro}, Claude Opus 5~\citep{anthropic2026claudeopus5}, and DeepSeek-V4-Pro~\citep{deepseek2026v4pro}. All prompting models receive the problem description, reference answer, and student submission. The single-error prompt and 4,865-item test set are unchanged across sets of prompted models. We report accuracy and macro F1; macro F1 is primary because the 14 classes are highly imbalanced. Finetuning details and complete per-class results appear in Appendices~\ref{sec:experiment-setup} and~\ref{sec:appendix-finetune-full}.

\label{sec:single-error-prompting}
For the multi-error task, models identify explicit errors and then apply a fixed counterfactual rule: minimally repair the explicit errors and add \texttt{Logic Error} only if the repaired program still violates a concrete requirement. Predictions are parsed as unordered label sets. We report exact-set match, sample-averaged F1, micro F1, and macro F1 over the 13 labels present in the expert subset. We retain the earlier \emph{contains} score only as a legacy diagnostic because it checks one normalized gold label rather than the complete expert set.

Provider APIs do not expose identical controls. DeepSeek and Gemini runs used temperature 0; GPT-5.5 rejected that setting and used the provider default; Claude Opus 5 rejected sampling parameters and used adaptive reasoning at medium effort. We record the exact served model identifier when the provider exposes one, the prompt text and hash, token usage, parsing failures, and run date. These differences are treated as part of the reproducibility record rather than silently normalized.

\section{RQ1: How Well Do Models Diagnose Student Code under the First-Outcome Protocol?}
\label{sec:results}
\label{sec:main-benchmark}
RQ1 establishes performance under the benchmark's original first-outcome construct. Since the test set, gold labels, and single-error prompt are unchanged, this section can compare the earlier and recent set of prompted models directly and ask whether the original model-selection conclusion still holds.
\subsection{Finetuning and Prompting Results}
\label{sec:single-error-finetuning-classification}
\label{sec:single-error-prompting-classification}

Table~\ref{tab:model-performance} summarizes Task C performance. The original finetuning comparison remains unchanged: CodeLlama-7B outperforms CodeBERT and reaches 80.6\% macro F1. The conclusion changes when the recent prompted models are included. All four exceed CodeLlama-7B, with macro F1 between 87.5\% and 93.8\%. GPT-5.5 achieves the highest macro F1, while accuracy and macro F1 give slightly different middle rankings.

\begin{table}[t]
\centering
\small
\begin{tabular}{llrr}
\hline
\textbf{Setting} & \textbf{Model} & \textbf{Accuracy} & \textbf{Macro F1} \\
\hline
Finetuned & CodeBERT & 73.4 & 45.2 \\
Finetuned & CodeLlama-7B & 91.7 & 80.6 \\
\hline
Earlier prompt & GPT-3.5 & 40.3 & 8.8 \\
Earlier prompt & GPT-4o & 71.5 & 28.0 \\
Earlier prompt & DeepSeek-V3 & 73.6 & 29.1 \\
Earlier prompt & Gemini 2.5 Pro & 85.9 & 71.9 \\
\hline
Recent prompt & DeepSeek-V4-Pro & 91.3 & 87.5 \\
Recent prompt & Claude Opus 5 & 93.3 & 90.8 \\
Recent prompt & Gemini 3.1 Pro Preview & 92.1 & 91.2 \\
Recent prompt & GPT-5.5 & \textbf{93.8} & \textbf{93.8} \\
\hline
\end{tabular}
\caption{Single-error Task C results on the 4,865-item test set (\%). The same first-outcome labels and single-error prompt are used for both sets of prompted models.}
\label{tab:model-performance}
\end{table}

The earlier result that a finetuned smaller model outperformed prompted LLMs was correct for the evaluated set of prompted models, but it does not persist across model generations. The benchmark now supports the opposite model-selection decision. This change is not caused by a new split or prompt; it comes from updating the evaluated models. The reversal itself is expected given the gap between the two model generations; the point is that a model-selection conclusion stated as a property of finetuning versus prompting did not survive one generation. Benchmark reports should therefore state model versions and evaluation dates and avoid presenting generation-specific rankings as durable properties of a method family.

Both finetuned models lose macro F1 as the taxonomy becomes finer (Table~\ref{tab:hier-summary}), and the earlier prompted models were far weaker on low-support classes than on common ones; complete Task A--C and per-class tables appear in Appendix~\ref{sec:appendix-finetune-full}. The recent models change the model-family comparison; they do not remove the class imbalance.

\subsection{Logic Error Overprediction}
\label{sec:logic-error-bias}

We retain the original definition of Logic Error Overprediction Rate, $\mathrm{LER}_{\mathrm{over}}$. For each non-Logic ground-truth class, we compute the proportion of instances predicted as \texttt{Logic Error}, then macro-average over the 13 classes. This class-balanced rate is useful because a small number of rare exceptions would otherwise have little influence on an aggregate error count.

\begin{table}[t]
\centering
\small
\begin{tabular}{llr}
\hline
\textbf{Set} & \textbf{Model} & $\mathrm{LER}_{\mathrm{over}}$ (\%) \\
\hline
Earlier & GPT-3.5 & 92.80 \\
Earlier & GPT-4o & 57.20 \\
Earlier & DeepSeek-V3 & 51.50 \\
Earlier & Gemini 2.5 Pro & 17.60 \\
\hline
Recent & DeepSeek-V4-Pro & 9.89 \\
Recent & Claude Opus 5 & 6.38 \\
Recent & Gemini 3.1 Pro Preview & 4.79 \\
Recent & GPT-5.5 & \textbf{4.18} \\
\hline
\end{tabular}
\caption{Logic Error Overprediction Rate under single-error prompting. Each value macro-averages the row-normalized \texttt{Logic Error} prediction rate over the 13 non-Logic gold classes.}
\label{tab:logic-overprediction-rate}
\end{table}

The earlier prompted models range from 17.6\% to 92.8\%, whereas the recent models range from 4.2\% to 9.9\% (Table~\ref{tab:logic-overprediction-rate}). Logic Error overprediction remains measurable under the standard protocol, but it is no longer the dominant failure mode suggested by the earlier set of prompted models. Most remaining errors are concentrated in \texttt{No Error}$\rightarrow$\texttt{Logic Error} and a few low-support classes. Detailed confusion matrices and per-class results are provided in Appendix~\ref{sec:appendix-single-prompting}.

\section{RQ2: Are Predictions Beyond the First Outcome Supported Along a Repair Path?}
\label{sec:multi-error-subset-construction}
RQ2 tests whether predictions that disagree with the first-outcome label are
supported by errors exposed along an expert repair path. We construct a targeted audit set, measure how often the original CodeBERT prediction appears in the expert set, and then examine whether predictive entropy helps locate unresolved disagreements.
\subsection{Locating Difficult Student Programs}

The diagnostic subset focuses on cases that are challenging for CodeBERT under the single-label Task C setting. We use two signals to identify such cases: model misclassification and high predictive entropy under the first-outcome evaluation. This targeted sampling allows us to examine whether these challenging predictions may reflect additional errors beyond the recorded execution label.

For the misclassification-based sample, we draw 40 misclassified instances from off-diagonal cells of the confusion matrix, excluding cells whose gold label is \texttt{No Error} or \texttt{Logic Error}. For the high-entropy sample, we take the 60 predictions with the highest entropy. Three submissions appear in both groups, leaving 97 unique cases. Fifteen expert annotators reviewed the cases over three rounds of verification. The resulting subset covers 13 error types and contains 187 labels, or 1.93 labels per submission on average.

\subsection{Expert Annotation Coverage}
\label{sec:multi-error-diagnostic}

We first ask how often a CodeBERT single-error prediction matches any label found by expert repair and re-execution. A prediction is covered when $P_i \cap G_i \neq \emptyset$, where $P_i$ is the model's singleton prediction and $G_i$ is the expert label set. The coverage rate is
\begin{equation}
    \mathrm{Coverage}=\frac{1}{N}\sum_{i=1}^{N}\mathds{1}[P_i\cap G_i\neq\emptyset].
\end{equation}

\begin{figure}[t]
    \centering
    \includegraphics[width=\linewidth]{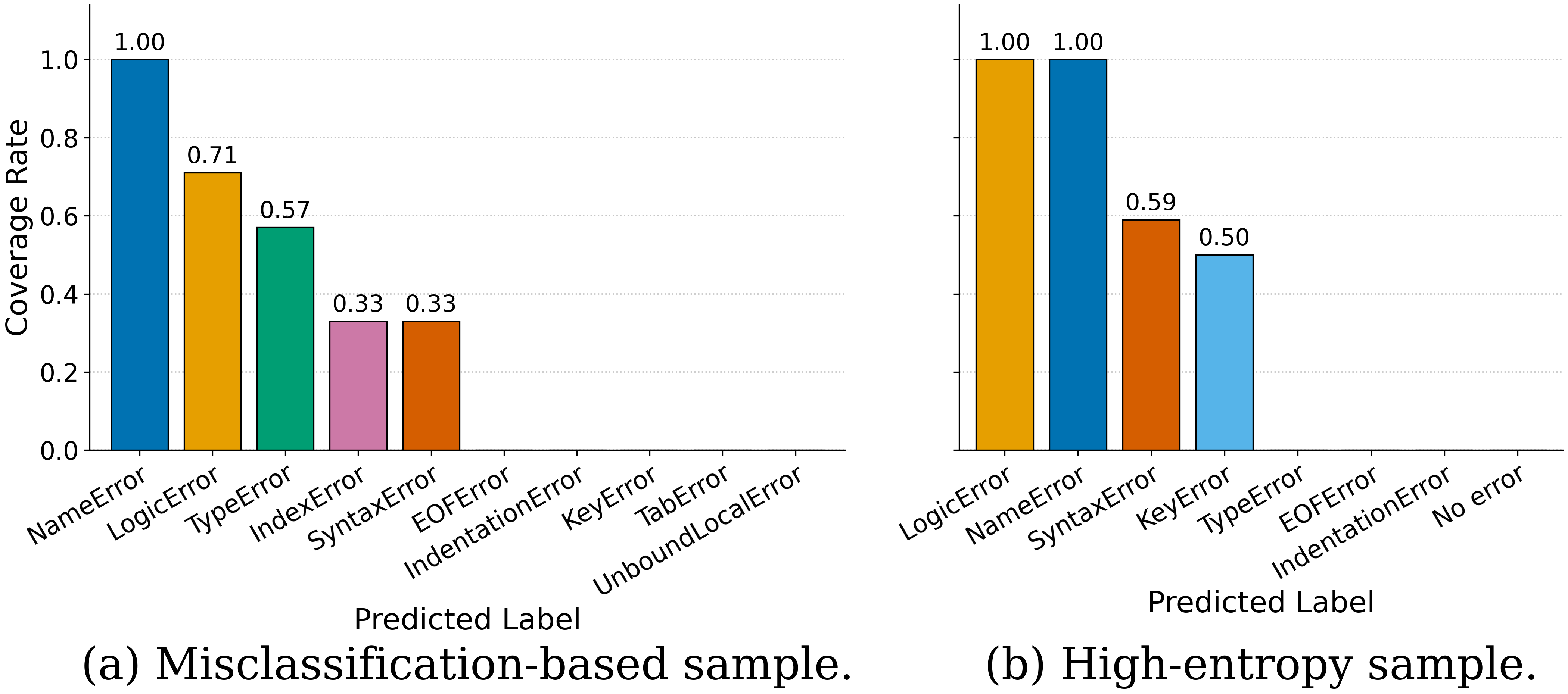}
    \caption{Coverage of CodeBERT's single-error predictions by the expert repair-conditional label sets. Coverage is reported by predicted error type for the two targeted sampling strategies: (a) misclassification-based sample; (b) high-entropy sample.}
    \label{fig:coverage-combined}
\end{figure}

Coverage remains uneven. In the misclassification-based sample, 50\% of predictions fall within the expert label set; in the high-entropy sample, 52\% do. \texttt{Name Error} reaches full coverage in both analyses, and \texttt{Logic Error} is frequently covered, while several rare types are not captured (Figure~\ref{fig:coverage-combined}). The repair-conditional sets therefore support some predictions that are scored as
wrong under first-outcome evaluation, but not the other half.

\subsection{Uncertainty as an Audit Signal}

\begin{figure}[t]
    \centering
    \includegraphics[width=0.50\linewidth]
    {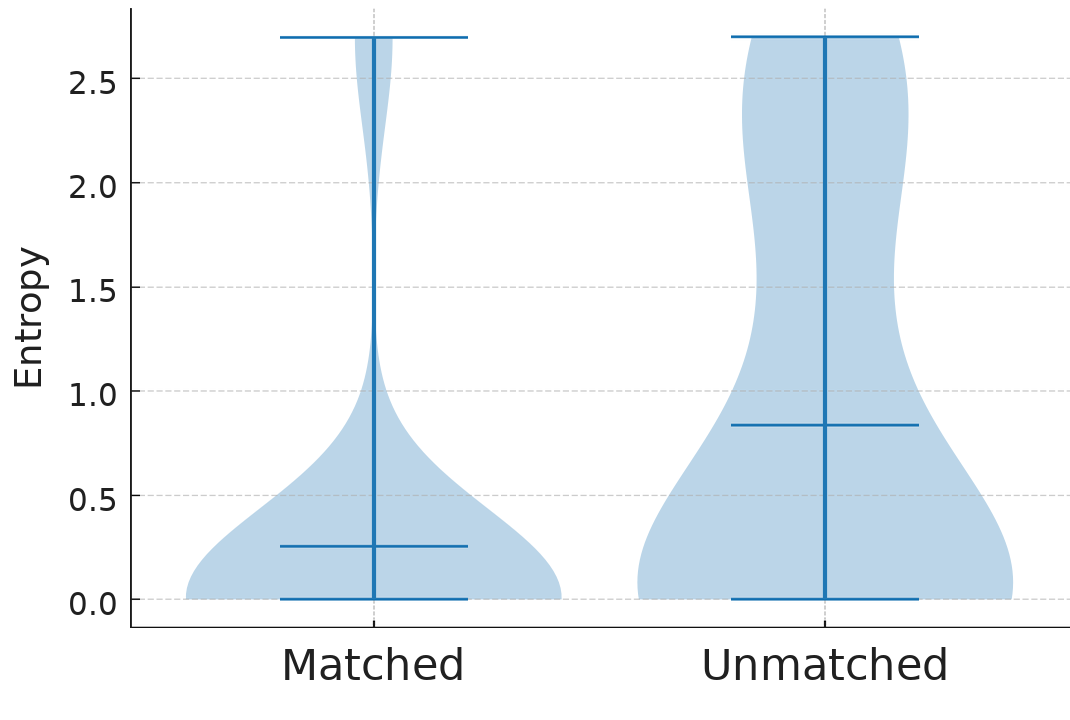}
    \caption{Prediction entropy for matched and unmatched cases in the diagnostic subset. A case is matched when CodeBERT's top-1 label appears in the expert repair-conditional label set.}
    \label{fig:entropy-violin}
\end{figure}

CodeBERT entropy is lower for predictions that match the expert set than for those that do not. Matched cases have mean entropy 0.25, compared with 0.84 for unmatched cases (Figure~\ref{fig:entropy-violin}). Both groups include low-entropy examples, so entropy cannot determine whether an individual
disagreement will be supported along a repair path. The separation appears mainly in the high-entropy region, where unmatched predictions are more common. Because high entropy was one of the two signals used to select the subset, this comparison describes the selected cases and does not estimate how entropy relates to disagreement in the full dataset. Entropy can help prioritize cases for review, but it does not determine whether a particular disagreement will appear on a repair path.

\section{RQ3: What Changes When Models Can Predict Multiple Errors?}
\label{sec:multi-error-prompting-classification}
RQ2 shows that a singleton prediction may match an error exposed later in an expert repair path. RQ3 turns that observation into a set-prediction task: models must recover the complete repair-conditional label set. We examine both full-set recovery and the partial-credit metrics that can make the same outputs appear more favorable.
\subsection{Set-Level Multi-Error Performance}

Table~\ref{tab:multi-set-results} evaluates the recent models against complete expert label sets. Exact-set match requires equality between the predicted and gold sets. Sample F1 averages per-submission set F1; micro F1 pools label decisions; macro F1 gives equal weight to the 13 labels present in the subset.
The legacy \emph{contains} score answers a narrower question: whether one normalized gold label appears anywhere in the output. It does not test whether the model recovered the complete repair-conditional set, so we do not use it for the current RQ3 claim. Earlier results produced with that score are retained only as a record in the appendix.

\begin{table}[t]
\centering
\small
\begin{tabular}{lrrrr}
\hline
\textbf{Model} & \textbf{Exact} & \textbf{Sample F1} & \textbf{Micro F1} & \textbf{Macro F1} \\
\hline
GPT-5.5 & 47.42 & 80.84 & \textbf{80.94} & \textbf{70.22} \\
Gemini 3.1 Pro Preview & \textbf{48.45} & \textbf{81.06} & 80.51 & 67.37 \\
Claude Opus 5 & 43.30 & 79.71 & 78.95 & 64.06 \\
DeepSeek-V4-Pro & 44.33 & 78.73 & 78.44 & 64.06 \\
\hline
\end{tabular}
\caption{Multi-error set metrics on the 97-item expert subset (\%). Macro F1 is computed over the 13 labels with gold support.}
\label{tab:multi-set-results}
\end{table}

No model identifies the complete set reliably: exact-set match is below 50\% for all four models. The ranking also depends on the metric. Gemini 3.1 Pro Preview has the highest exact match and sample F1, while GPT-5.5 has the highest micro and macro F1. The exact-match 95\% intervals overlap, spanning approximately 38--59\% for Gemini and 38--58\% for GPT-5.5, so the subset does not support a claim that one model is decisively best.

\subsection{The Evaluation Configuration Reveals Secondary Logic Predictions}

Set-level scores do not show which extra labels models add. We therefore
inspect the current multi-error outputs for the 45 audited submissions whose
expert sets contain no \texttt{Logic Error}. The four models add an
unsupported \texttt{Logic Error} label to 21--26 of these predictions
(Table~\ref{tab:paired-logic}). In all but four Gemini 3.1 Pro Preview
outputs, the label is not listed first; it appears alongside an
explicit-error diagnosis.
Under the single-error protocol, the same models predict \texttt{Logic
Error} on none of these 45 submissions (Table~\ref{tab:paired-logic},
first column), consistent with their low overall overprediction rates
(4.2--9.9\%, Section~\ref{sec:logic-error-bias}). Given a multi-label output
space, they attach the label to roughly half. Although the models obtain
78.7--81.1\% sample F1, this systematic secondary false-positive pattern is
one reason exact-set recovery stays below 50\%: strong partial-match
performance and a persistent label bias coexist.

\begin{table}[t]
\centering
\small
\begin{tabular}{lrrr}
\hline
\textbf{Model} & \textbf{Single-error LE} & \textbf{Any-position LE} & \textbf{First-listed LE} \\
\hline
GPT-5.5 & 0/45 & 21/45 (46.7\%) & 0/45 \\
Gemini 3.1 Pro Preview & 0/45 & 24/45 (53.3\%) & 4/45 \\
Claude Opus 5 & 0/45 & 22/45 (48.9\%) & 0/45 \\
DeepSeek-V4-Pro & 0/45 & 26/45 (57.8\%) & 0/45 \\
\hline
\end{tabular}
\caption{Logic Error overprediction on the same 45 audited submissions
without expert-annotated Logic Error. Single-error LE counts the label
under single-error prompting; the multi-error columns count it anywhere
in the predicted set (any-position) and in first position (first-listed).}
\label{tab:paired-logic}
\end{table}

The single-label evaluation therefore makes the \texttt{Logic Error} tendency appear absent on these audited cases, while the multi-label output space exposes it as secondary co-prediction. The two configurations change the prompt, the allowed output cardinality, and the scoring rule together; in particular, the multi-error prompt adds an explicit repair-then-check instruction. Our comparison therefore does not isolate any one of these components as the cause of the difference.

\section{Discussion and Limitations}

The first-outcome protocol is appropriate when the task is to reproduce the first outcome shown by an Online Judge or IDE. It is not an exhaustive account of a student program. This distinction matters most in the cases on which model and system feedback disagree. Some disagreements are ordinary model errors; others arise because the model names an error that becomes visible only after an earlier fault is repaired. Reporting the benchmark as ``code error diagnosis'' without specifying this boundary makes those cases difficult to interpret.

Model explanations from the earlier single-error experiments help clarify why
execution-grounded and intent-grounded diagnoses diverge. One recurring
pattern is intent-first reasoning: in one example, a full-width colon causes
an immediate parse-time failure, yet the model evaluates the ordering of the
leap-year conditions and predicts \texttt{Logic Error}. The reverse also
occurs: models treat unconventional spacing or naming as evidence of a syntax
error even though the code runs and fails only on its output. Both patterns
show why a label defined by interpreter order can disagree with a model that
reads the program as a whole; additional examples appear in
Appendix~\ref{appendix:behavioral-alignment}.

The updated prompting results expose a second source of instability. The earlier comparison favored a finetuned smaller model; the recent comparison favors prompted models, even though the test set and single-error prompt are unchanged. Within the multi-error task, the metric also affects the result: Gemini 3.1 Pro Preview has the highest exact-set match and sample F1, while GPT-5.5 has the highest micro and macro F1. A permissive membership test gives a far more favorable result than exact-set scoring because it does not require recovery of the complete label set. Model identifiers, evaluation dates, prompt provenance, uncertainty, and set-appropriate metrics are therefore part of the claim, not merely implementation details.

The diagnostic subset should be read within its design. The 97 submissions were selected using CodeBERT errors and entropy in order to expose cases where a first-outcome label might be incomplete. They demonstrate that such failures exist and show how they are structured, but they cannot estimate their prevalence among all 48,646 submissions. The subset is also small for rare-label comparisons, and the bootstrap intervals do not reliably separate the recent models. A random annotated sample, together with pre-consensus agreement statistics, would support broader population-level claims. We did not record annotator agreement before adjudication, and each repair-conditional set depends on the repair the annotator chose: a different fix for an early error could expose different downstream errors. Independent double annotation of the same subset would quantify both effects.

A further caveat concerns data exposure. \datasetname was publicly released before three of the four recent models finished training; only the pinned GPT-5.5 snapshot predates the release. We therefore cannot rule out that part of the single-error improvement reflects exposure to the benchmark rather than capability. The audit results are informative under this uncertainty: even if first-outcome scores were inflated by exposure, exact-set match stays below 50\% and secondary \texttt{Logic Error} predictions persist on roughly half of the audited submissions. Scores obtained under a permissive protocol can rise for reasons unrelated to the construct being measured, while a stricter protocol continues to expose the gap.

Several limitations come from the dataset and model APIs. \datasetname contains only Python submissions, and its problem descriptions are primarily Chinese; the present experiments do not establish transfer across natural languages, programming languages, curricula, or Online Judge configurations. Provider APIs do not expose identical sampling and reasoning controls, so model settings cannot be perfectly matched. The dataset also contains no student demographic information, which precludes subgroup analysis. All available identifiers were irreversibly anonymized before analysis.

\section{Conclusion}

We used \datasetname to examine what first-outcome labels can and cannot support in evaluations of student-code diagnosis. Under the standard single-error protocol, the four recent prompted models all outperform the strongest finetuned baseline in macro F1. The earlier finding that finetuning a smaller model was preferable to prompting a larger one was accurate for the models tested at that time, but it is not stable across model generations.

The expert audit addresses a different question. On 97 deliberately difficult submissions, an expert repair path often exposes labels beyond the first system outcome. The expert repair-conditional sets cover roughly half of CodeBERT's predictions in both targeted samples. In the misclassification-based sample, this means that half of the predictions scored as wrong under first-outcome evaluation are supported along the repair path. Permissive scoring masks the remaining difficulty: exact-set match is below 50\% for every recent model, and the highest-ranked model changes with the metric.

The comparison of single- and multi-error outputs also reveals a behavior hidden by the one-label task. On the 45 audited submissions without expert-annotated \texttt{Logic Error}, the recent models never choose that label when limited to a single prediction, but frequently add it as a secondary label when multiple outputs are allowed. The two configurations expose different parts of model behavior.

Execution-derived labels remain valuable because they are reproducible and closely tied to real system feedback. Their value depends, however, on stating the construct precisely. Future work should pair execution-labeled datasets with randomly sampled expert audits, report complete-set metrics alongside per-label ones, and preserve model and prompt provenance.

\section*{Acknowledgments}
We are grateful to the volunteer expert annotators whose careful work made the multi-error diagnostic subset possible. We also thank the TAE reviewers for their constructive feedback.

\bibliographystyle{plainnat}
\bibliography{custom}
\clearpage

\appendix
\section{Supplementary Material}
\label{sec:appendix}

\subsection{Task C Multi-class Classification}
\label{sec:appendix-taxonomy-defs}

\begin{description}
  \item[\texttt{Logic Error}:] Code that compiles successfully but fails one or more test cases due to incorrect logic or implementation.
  \item[\texttt{Syntax Error}:] Raised when the parser encounters a syntax error in the code.
  \item[\texttt{Name Error}:] Raised when a local or global name is not found.
  \item[\texttt{Type Error}:] Raised when an operation or function is applied to an object of inappropriate type.
  \item[\texttt{Indentation Error}:] Raised when there is incorrect indentation in the code.
  \item[\texttt{Unbound Local Error}:] Raised when a local variable is referenced before it has been assigned.
  \item[\texttt{Key Error}:] Raised when a dictionary key is not found.
  \item[\texttt{Index Error}:] Raised when a sequence subscript is out of range.
  \item[\texttt{EOF Error}:] Raised when the input() function hits an end-of-file condition (EOF) without reading any data.
  \item[\texttt{Recursion Error}:] Raised when the maximum recursion depth is exceeded.
  \item[\texttt{Value Error}:] Raised when a function receives an argument of the correct type but an inappropriate value.
  \item[\texttt{Tab Error}:] Raised when indentation consists of inconsistent use of tabs and spaces.
  \item[Attribute Error:] Raised when an attribute reference or assignment fails.
  \item[Runtime Error:] Raised when an error is detected that does not fall in any of the other categories.
  \item[Syntax Warning:] Raised for dubious syntactic features but not necessarily invalid syntax.
  \item[Zero Division Error:] Raised when division or modulo by zero takes place for all numeric types.
  \item[Memory Error:] Raised when an operation runs out of memory.
  \item[Module Not Found Error:] Raised when a module could not be found.
  \item[\texttt{No Error}:] Code that compiles and passes all test cases successfully.

\end{description}

\begin{table}[htbp]
\centering
\small
\begin{tabular}{p{0.40\linewidth}lll}
\hline
\textbf{Raw annotation label} & \textbf{Task C (14)} & \textbf{Task B (3)} & \textbf{Task A (2)} \\
\hline
No Error & No Error & No Error & No Error \\
Logic Error & Logic Error & Logic Error & Error \\
Syntax Error & Syntax Error & Explicit Error & Error \\
Name Error & Name Error & Explicit Error & Error \\
Type Error & Type Error & Explicit Error & Error \\
Indentation Error & Indentation Error & Explicit Error & Error \\
Unbound Local Error & Unbound Local Error & Explicit Error & Error \\
Key Error & Key Error & Explicit Error & Error \\
Index Error & Index Error & Explicit Error & Error \\
EOF Error & EOF Error & Explicit Error & Error \\
Recursion Error & Recursion Error & Explicit Error & Error \\
Value Error & Value Error & Explicit Error & Error \\
Tab Error & Tab Error & Explicit Error & Error \\
Attribute Error, Runtime Error, Syntax Warning, Zero Division Error, Memory Error, Module Not Found Error & Other Errors & Explicit Error & Error \\
\hline
\end{tabular}
\caption{Mapping from the 18 raw error types plus \texttt{No Error} used in annotation (Appendix~\ref{sec:annotation-guidelines}) to the three evaluation levels. The six low-frequency types in the last row are merged into \texttt{Other Errors} for Task C, giving 14 classes.}
\label{tab:taxonomy-mapping}
\end{table}

\subsection{Motivation for Our Taxonomy Design}
\label{sec:appendix-taxonomy-motivation}
This section explains the considerations that motivated the design of our taxonomy.

At present, there is no consensus on taxonomies for single-label or multi-label code error classification.
For example, the AOJ dataset aligns error categories with online judge outcomes, such as Runtime Error and Wrong Answer.
COJ2022~\citep{10.1145/3539618.3591680} computes line level diffs between student code submissions with potential errors and their corrected versions using difflib, and labels errors such as Function, Declaration, etc. The taxonomy in \citet{dubniczky2025castlebenchmarkingdatasetstatic} includes 25 types of CWE vulnerabilities, such as buffer overflow, SQL injection, cross-site scripting, and others. Other work \citep{Shirafuji_2023} applies AST based analysis and reference solutions to the AOJ dataset for line level error judgments, yielding 21 coarse grained classes and 55 fine grained categories, including output, input, variable conversion, etc.

Current error taxonomies in code mostly focus on algorithmic optimization or are framed purely from a human-centered perspective.
Many line-wise mismatched code categories are essentially code optimizations rather than actual errors.
Such categories would potentially cause confusion for both human learners and automated systems, and they may not always be helpful for detecting real errors.
Meanwhile, we observe an increasing number of studies that integrate LLMs into IDEs to assist with code generation and error analysis.
Some work examines the debugging challenges of LLM-powered IDE tools in code understanding, generation, and automated repair, and presents an automated testing system that enables error detection in lockstep with codebase updates~\citep{lee2024improvingllmclassificationlogical,Wang:2025}.
Others~\citep{chan2024copilotevaluationharness} explore Copilot's iterative error-checking mechanisms following code edits.
These developments show why an error taxonomy should target real errors and remain interpretable to both IDEs and human developers.

In response, we propose a three-level, process-based code error taxonomy that uses Python's official error types as the foundational layer to locate genuine error categories, serving both human coding education and IDE functionalities.

\subsection{Source Data and Platform}
\label{sec:source-data-platform}
\begin{figure}[htbp]
    \centering
    \setlength{\abovecaptionskip}{4pt}
    \setlength{\belowcaptionskip}{4pt}
    \vspace{-0.8em}  %
    
    \begin{minipage}{\linewidth}
        \centering
        \includegraphics[width=0.92\linewidth]{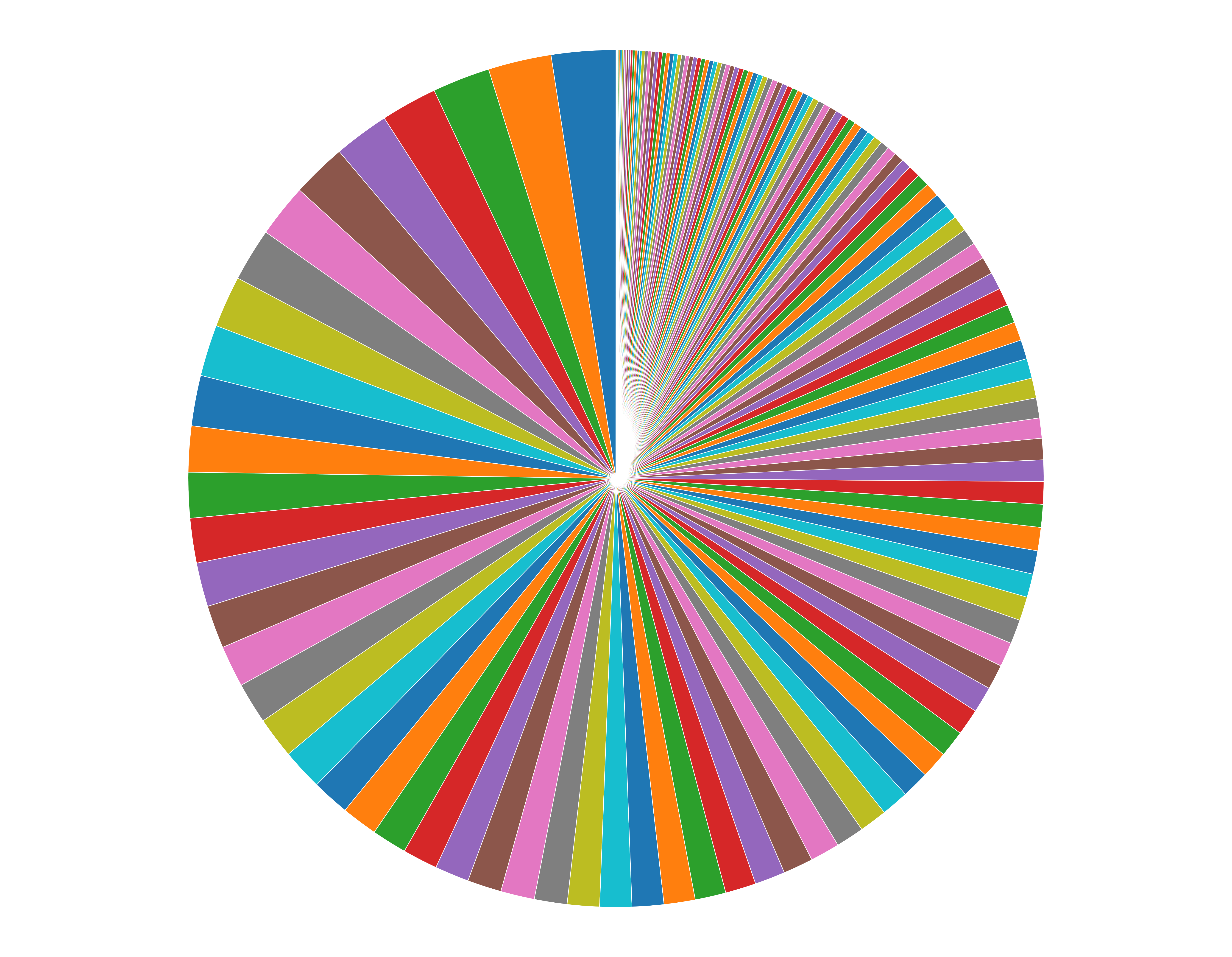}
        \caption{Question ID Distribution.}
        \label{fig:question-id-dist}
    \end{minipage}
    
    \vspace{0.4em}  %
    
    \begin{minipage}{\linewidth}
        \centering
        \includegraphics[width=0.92\linewidth]{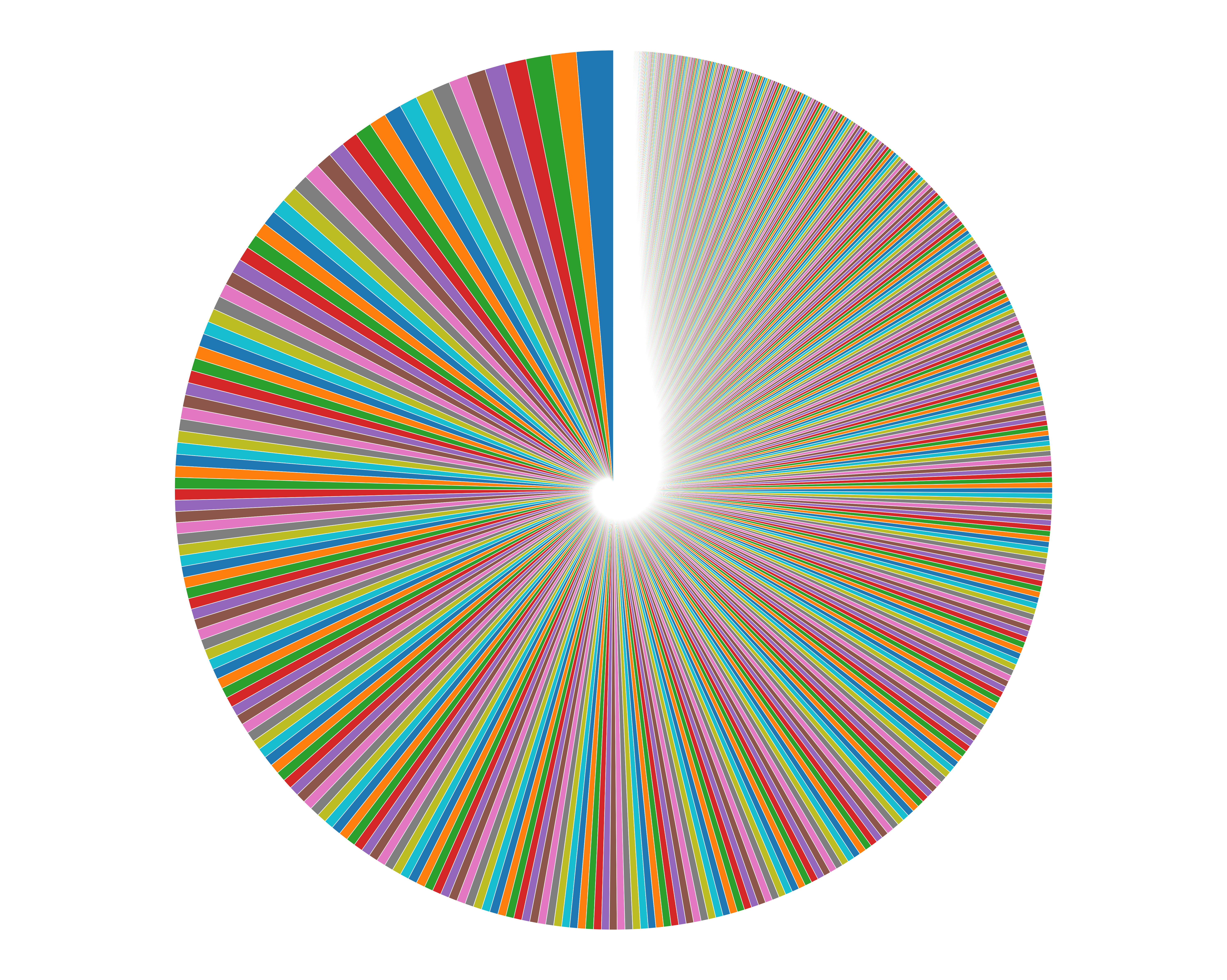}
        \caption{User Distribution.}
        \label{fig:user-id-dist}
    \end{minipage}
    
    \vspace{-0.8em}  %
\end{figure}

We describe the sources, collection process, and ethical considerations of the raw data used in this section.
Our data originate from an existing online programming education platform developed and maintained by Circle Cat's engineering team. The platform provides Python and Java programming exercises, allowing learners to submit code solutions that are automatically compiled, executed, and evaluated through an online judge system, with immediate feedback returned to learners.
The dataset was collected from historical platform logs and does not involve direct recruitment, paid participation, or experimental intervention. All code submissions were generated organically during normal platform usage. Data collection and usage follow the platform’s terms of use, which inform users that anonymized data may be used for educational and research purposes.
From the code runner backend, we extracted 48,646 complete and valid code submission samples from 153 distinct questions and 579 users with different background as the initial dataset. The Question and User distribution are shown in Figure~\ref{fig:question-id-dist} and Figure~\ref{fig:user-id-dist}. These code submissions were cleaned and post-processed before expert annotation to construct a fine-grained multi-error code error dataset based on real learner-written code.
Because the data were obtained from pre-existing educational records and involve no human subject intervention, ethics review board approval was not required under institutional guidelines. Details regarding data anonymization, privacy considerations, and annotation procedures are provided in Section~\ref{sec:dataset-preprocess}.

\subsection{Dataset Preprocess and Annotation}
\label{sec:dataset-preprocess}
The dataset used in this study is derived from the Moodle-based educational platform described above, which provides anonymized interaction logs (e.g., attempt IDs and attemptstep IDs) for research purposes. All identifiers were irreversibly anonymized before analysis, and no directly identifying information is included in the released data.
Annotation for the code error taxonomy was performed by a team of 15 volunteer researchers familiar with Python programming and educational assessment. The 15 annotators consisted of software engineers and researchers with backgrounds in computer science and educational assessment. All expert annotators had 3--5 years of engineering experience in Python language and were familiar with common student programming errors at the introductory level. The annotation guidelines and examples are provided below.

\subsubsection{Annotation Guidelines}
\label{sec:annotation-guidelines}
Annotators were provided with a detailed error taxonomy comprising 18 error types plus a ``no error'' category (ID 0). Each error type is defined by its associated runtime or compile-time behavior: \texttt{Logic Error} (ID 1) applies when code executes without explicit exceptions but fails one or more test cases; all other error types (IDs 2--18) are identified by matching the execution output against a specific error string (e.g., \texttt{Syntax Error}, \texttt{Name Error}, \texttt{Type Error}).

For each student code sample identified by a (\textit{Question ID}, \textit{Attempt ID}) pair, annotators followed the procedure below:

\begin{enumerate}
    \item Locate the specific programming problem on the online judge using the Question ID.
    \item Paste the student's answer into the judge, observe the first error, and increment the corresponding error-type count in the annotation spreadsheet.
    \item Fix \emph{only} that single error (referring to the expected answer as a reference), and verify that re-execution no longer produces the same error at the same location.
    \item Repeat steps 2--3 until no explicit runtime or compile-time errors remain.
    \item If the problem includes test cases, run them: failure to pass all test cases indicates a \texttt{Logic Error}; passing all test cases marks the sample as complete.
\end{enumerate}

\subsection{Experiment Setup Details}
\label{sec:experiment-setup}
We conduct finetuning experiments using CodeLlama-7B as the base model. The task is formulated as a multi-class classification problem with 14 error categories. The key hyperparameters required for reproducibility are summarized below.

\begin{itemize}
  \item Base model: CodeLlama-7B
  \item Number of labels: 14
  \item Maximum sequence length: 512
  \item Training epochs: 5
  \item Effective batch size: 12 (batch size 4 with gradient accumulation steps 3)
  \item Learning rate: $2 \times 10^{-4}$
  \item Weight decay: 0.01
  \item Random seed: 88
\end{itemize}

We apply LoRA-based parameter-efficient finetuning with the following configuration: LoRA rank $r=16$, LoRA alpha $=32$, and dropout $=0.05$. To improve memory efficiency, we employ 4-bit quantization using the NF4 scheme with bfloat16 computation. Training is performed using the AdamW optimizer with learning rate warmup and early stopping based on validation loss. Gradient checkpointing is enabled to reduce memory usage. Other training and implementation details follow standard practice.

For API-based models, we use prompt-based inference without parameter updates. The earlier set of prompted models contains GPT-3.5-Turbo, GPT-4o, DeepSeek-V3, and Gemini 2.5 Pro; the recent set of prompted models contains GPT-5.5, Gemini 3.1 Pro Preview, Claude Opus 5, and DeepSeek-V4-Pro. The single-error prompt is unchanged between the two sets of models. DeepSeek and Gemini runs use temperature 0. GPT-5.5 rejects that parameter and uses the provider default, while Claude Opus 5 rejects sampling parameters and uses adaptive reasoning at medium effort. Random seed control is not available for these APIs.

All sample extraction procedures are implemented using deterministic scripts with fixed random seeds to ensure reproducibility. Prior to evaluation, we apply rule-based sampling and filtering to control the distribution of extracted prediction samples.

\subsection{Single-Error Finetuning Classification}
\begin{table}[htbp]
\centering
\small
\begin{tabular}{lccc}
\hline
\textbf{Model} & \textbf{Task A} & \textbf{Task B} & \textbf{Task C} \\
\hline
CodeBERT & 80.4 & 72.9 & 45.2 \\
CodeLlama-7B & 96.3 & 93.5 & 80.6 \\
\hline
\end{tabular}
\caption{Macro F1 (\%) of the finetuned models across the three taxonomy levels on the test set (Task A: binary, Task B: three-class, Task C: 14-class). Complete per-class tables appear in Appendix~\ref{sec:appendix-finetune-full}.}
\label{tab:hier-summary}
\end{table}

\subsubsection{Test Set Results (Complete Tables)}
\label{sec:appendix-finetune-full}
The following tables report the complete per-class test-set results for CodeBERT and CodeLlama-7B on Tasks A, B, and C under the problem-level \texttt{QuestionID} split; Table~\ref{tab:hier-summary} summarizes their macro F1.

\begin{table}[!t]
  \centering
  \resizebox{0.8\columnwidth}{!}{
      \begin{tabular}{lccc}
        \hline
        \textbf{Class} & \textbf{Precision (\%)} & \textbf{Recall (\%)} & \textbf{F1-score (\%)} \\
        \hline
        \texttt{No Error} & 64.2 & 91.7 & 75.5 \\
        Error     & 95.4 & 77.0 & 85.2 \\
        \hline
        Macro avg     & 79.8 & 84.4 & 80.4 \\
        Weighted avg  & 85.7 & 81.6 & 82.2 \\
        \hline
      \end{tabular}
    }
  \caption{\label{tab:codebert-taskA}
    CodeBERT Task A (Binary Classification) results on the test set under problem-level split by \texttt{QuestionID}. The following tables use the same split.
  }
\end{table}

\begin{table}[t]
  \centering
  \resizebox{0.8\columnwidth}{!}{
      \begin{tabular}{lccc}
        \hline
        \textbf{Class} & \textbf{Precision (\%)} & \textbf{Recall (\%)} & \textbf{F1-score (\%)} \\
        \hline
        \texttt{No Error} & 68.2 & 91.1 & 78.0 \\
        Explicit Error & 84.7 & 51.7 & 64.2 \\
        \texttt{Logic Error} & 74.6 & 78.4 & 76.5 \\
        \hline
        Macro avg     & 75.8 & 73.7 & 72.9 \\
        Weighted avg  & 75.8 & 73.9 & 73.1 \\
        \hline
      \end{tabular}
    }
  \caption{\label{tab:codebert-taskB}
    CodeBERT Task B (Three-class Classification) results on the test set.
  }
\end{table}

\begin{table}[!t]
  \centering
  \resizebox{0.9\columnwidth}{!}{
  \begin{tabular}{lccc}
    \hline
    \textbf{Class} & \textbf{Precision (\%)} & \textbf{Recall (\%)} & \textbf{F1-score (\%)} \\
    \hline
    0~\texttt{No Error}              & 67 & 95 & 78 \\
    1~\texttt{Logic Error}            & 79 & 79 & 79 \\
    2~\texttt{Syntax Error}           & 92 & 71 & 80 \\
    3~\texttt{Name Error}             & 72 & 46 & 56 \\
    4~\texttt{Type Error}             & 41 & 19 & 26 \\
    5~\texttt{Indentation Error}      & 89 & 25 & 39 \\
    6~\texttt{Unbound Local Error}     & 47 & 18 & 26 \\
    7~\texttt{Key Error}              & 77 & 59 & 67 \\
    8~\texttt{Index Error}            & 30 & 21 & 25 \\
    9~\texttt{EOF Error}              & 75 & 40 & 52 \\
    10~\texttt{Recursion Error}       & 0 & 0 & 0 \\
    11~\texttt{Value Error}           & 68 & 29 & 41 \\
    12~\texttt{Tab Error}             & 56 & 77 & 65 \\
    13~\texttt{Other Errors}         & 0 & 0 & 0 \\
    \hline
    Macro avg     & 57.0 & 41.0 & 45.2 \\
    Weighted avg  & 72.9 & 73.4 & 71.0 \\
    \hline
  \end{tabular}
  }
  \caption{\label{tab:codebert-task3}
    CodeBERT Task C (Fine-grained Multi-class Classification) results on the test set. Class indices follow the taxonomy in Section~\ref{sec:taxonomy-design} and match the row/column order of the confusion matrices and the label IDs used in the prompts (Appendix~\ref{sec:appendix-prompt}). Per-class precision/recall/F1 at integer precision; Macro and Weighted averages at one decimal place, same for Table~\ref{tab:codellama-taskC}.
  }
  
\end{table}

\begin{table}[t]
  \centering
  \resizebox{0.8\columnwidth}{!}{
  \begin{tabular}{lccc}
    \hline
    \textbf{Class} & \textbf{Precision (\%)} & \textbf{Recall (\%)} & \textbf{F1-score (\%)} \\
    \hline
    \texttt{No Error} & 92.1 & 98.2 & 95.0 \\
    \texttt{Error}     & 99.2 & 96.2 & 97.7 \\
    \hline
    Macro avg     & 95.7 & 97.2 & 96.3 \\
    Weighted avg  & 97.0 & 96.8 & 96.8 \\
    \hline
  \end{tabular}
  }
  \caption{\label{tab:codellama-task1}
    CodeLlama Task A (Binary Classification) results on the test set.
  }
\end{table}

\begin{table}[t]
  \centering
  \resizebox{0.9\columnwidth}{!}{
  \begin{tabular}{lccc}
    \hline
    \textbf{Class} & \textbf{Precision (\%)} & \textbf{Recall (\%)} & \textbf{F1-score (\%)} \\
    \hline
    \texttt{No Error}        & 92.5 & 97.8 & 95.1 \\
    \texttt{Explicit Error}  & 92.2 & 93.9 & 93.1 \\
    \texttt{Logic Error}     & 95.3 & 89.3 & 92.2 \\
    \hline
    Macro avg     & 93.3 & 93.7 & 93.5 \\
    Weighted avg  & 93.5 & 93.4 & 93.4 \\
    \hline
  \end{tabular}
  }
  \caption{\label{tab:codellama-taskB}
    CodeLlama Task B (Three-class Classification) results on the test set.
  }
\end{table}

\begin{table}[!t]
  \centering
  \resizebox{0.9\columnwidth}{!}{
  \begin{tabular}{lccc}
    \hline
    \textbf{Class} & \textbf{Precision (\%)} & \textbf{Recall (\%)} & \textbf{F1-score (\%)} \\
    \hline
    0~\texttt{No Error}              & 90 & 99 & 94 \\
    1~\texttt{Logic Error}            & 96 & 89 & 93 \\
    2~\texttt{Syntax Error}           & 95 & 97 & 96 \\
    3~\texttt{Name Error}             & 81 & 93 & 86 \\
    4~\texttt{Type Error}             & 85 & 82 & 84 \\
    5~\texttt{Indentation Error}      & 97 & 91 & 94 \\
    6~\texttt{Unbound Local Error}     & 67 & 72 & 69 \\
    7~\texttt{Key Error}              & 88 & 87 & 88 \\
    8~\texttt{Index Error}            & 45 & 36 & 40 \\
    9~\texttt{EOF Error}              & 96 & 96 & 96 \\
    10~\texttt{Recursion Error}       & 100 & 60 & 75 \\
    11~\texttt{Value Error}           & 88 & 64 & 74 \\
    12~\texttt{Tab Error}             & 85 & 85 & 85 \\
    13~\texttt{Other Errors}         & 82 & 40 & 54 \\
    \hline
    Macro avg     & 85.0 & 78.0 & 80.6 \\
    Weighted avg  & 91.9 & 91.7 & 91.6 \\
    \hline
  \end{tabular}
  }
  \caption{\label{tab:codellama-taskC}
    CodeLlama Task C (Fine-grained Multi-class Classification) results on the test set.
  }
\end{table}

\subsubsection{Dev Set Results}
\label{sec:appendix-finetuning-test}

The following tables report dev set performance for all six finetuning experiments (CodeBERT and CodeLlama on Tasks A, B, and C) under the problem-level \texttt{QuestionID} split. These complement the test set results reported in the main text. For Task C, class~7 (\texttt{Key Error}) is absent from the dev set due to the random QuestionID split, but present in the test set.
 
\begin{table}[htbp]
  \centering
  \begin{tabular}{lccc}
    \hline
    \textbf{Class} & \textbf{Precision (\%)} & \textbf{Recall (\%)} & \textbf{F1-score (\%)} \\
    \hline
    \texttt{No Error} & 73.1 & 94.6 & 82.5 \\
    \texttt{Error}     & 95.9 & 78.1 & 86.1 \\
    \hline
    Macro avg     & 84.5 & 86.4 & 84.3 \\
    Weighted avg  & 87.1 & 84.5 & 84.7 \\
    \hline
  \end{tabular}
  \caption{
    CodeBERT Task A (Binary Classification) results on the dev set under problem-level split by \texttt{QuestionID}. Per-class precision/recall/F1 are reported at one decimal place; Macro and Weighted averages follow the stored full-precision computation.
  }
  \label{tab:appendix-codebert-taskA-test}
\end{table}
 
\begin{table}[htbp]
  \centering
  \begin{tabular}{lccc}
    \hline
    \textbf{Class} & \textbf{Precision (\%)} & \textbf{Recall (\%)} & \textbf{F1-score (\%)} \\
    \hline
    \texttt{No Error}        & 81.1 & 82.0 & 81.6 \\
    \texttt{Explicit Error}  & 80.3 & 71.8 & 75.8 \\
    \texttt{Logic Error}     & 67.1 & 73.2 & 70.1 \\
    \hline
    Macro avg     & 76.2 & 75.7 & 75.8 \\
    Weighted avg  & 76.6 & 76.2 & 76.3 \\
    \hline
  \end{tabular}
  \caption{
    CodeBERT Task B (Three-class Classification) results on the dev set under problem-level split by \texttt{QuestionID}. Per-class precision/recall/F1 are reported at one decimal place; Macro and Weighted averages follow the stored full-precision computation.
  }
  \label{tab:appendix-codebert-taskB-test}
\end{table}
 
\begin{table}[htbp]
  \centering
\begin{tabular}{lccc}
    \hline
    \textbf{Class} & \textbf{Precision (\%)} & \textbf{Recall (\%)} & \textbf{F1-score (\%)} \\
    \hline
    0~\texttt{No Error}              & 77 & 96 & 85 \\
    1~\texttt{Logic Error}            & 78 & 72 & 75 \\
    2~\texttt{Syntax Error}           & 93 & 83 & 87 \\
    3~\texttt{Name Error}             & 78 & 56 & 65 \\
    4~\texttt{Type Error}             & 69 & 36 & 47 \\
    5~\texttt{Indentation Error}      & 80 & 60 & 69 \\
    6~\texttt{Unbound Local Error}     & 67 & 19 & 30 \\
    7~\texttt{Key Error}              & --- & --- & --- \\
    8~\texttt{Index Error}            & 100 & 50 & 67 \\
    9~\texttt{EOF Error}              & 6 & 10 & 7 \\
    10~\texttt{Recursion Error}       & 0 & 0 & 0 \\
    11~\texttt{Value Error}           & 0 & 0 & 0 \\
    12~\texttt{Tab Error}             & 33 & 40 & 36 \\
    13~\texttt{Other Errors}         & 0 & 0 & 0 \\
    \hline
    Macro avg     & 52 & 40 & 44 \\
    Weighted avg  & 79 & 79 & 78 \\
    \hline
  \end{tabular}
  \caption{\label{tab:appendix-codebert-taskC-test}
    CodeBERT Task C (Fine-grained Multi-class Classification) results on the dev set under problem-level split by \texttt{QuestionID}. Class~7 (\texttt{Key Error}) shows ``---'' because the random \texttt{QuestionID}-based split yields no \texttt{Key Error} samples in the dev set. Per-class precision/recall/F1 are reported at decimal precision; Macro and Weighted averages follow the stored full-precision computation (one decimal place). Error types are indexed according to the taxonomy described in Section~\ref{sec:taxonomy-design}.
  }
\end{table}
 
\begin{table}[htbp]
  \centering
  \begin{tabular}{lccc}
    \hline
    \textbf{Class} & \textbf{Precision (\%)} & \textbf{Recall (\%)} & \textbf{F1-score (\%)} \\
    \hline
    \texttt{No Error} & 96.5 & 97.2 & 96.8 \\
    \texttt{Error}     & 98.3 & 97.8 & 98.0 \\
    \hline
    Macro avg     & 97.4 & 97.5 & 97.4 \\
    Weighted avg  & 97.6 & 97.6 & 97.6 \\
    \hline
  \end{tabular}
  \caption{\label{tab:appendix-codellama-taskA-test}
    CodeLlama Task A (Binary Classification) results on the dev set under problem-level split by \texttt{QuestionID}. Per-class precision/recall/F1 are reported at one decimal place; Macro and Weighted averages follow the stored full-precision computation.
  }
\end{table}
 
\begin{table}[htbp]
  \centering
  \begin{tabular}{lccc}
    \hline
    \textbf{Class} & \textbf{Precision (\%)} & \textbf{Recall (\%)} & \textbf{F1-score (\%)} \\
    \hline
    \texttt{No Error}        & 96.9 & 97.2 & 97.1 \\
    \texttt{Explicit Error}  & 95.6 & 97.7 & 96.6 \\
    \texttt{Logic Error}     & 94.9 & 92.4 & 93.6 \\
    \hline
    Macro avg     & 95.8 & 95.8 & 95.8 \\
    Weighted avg  & 95.9 & 95.9 & 95.9 \\
    \hline
  \end{tabular}
  \caption{\label{tab:appendix-codellama-taskB-test}
    CodeLlama Task B (Three-class Classification) results on the dev set under problem-level split by \texttt{QuestionID}. Per-class precision/recall/F1 are reported at one decimal place; Macro and Weighted averages follow the stored full-precision computation.
  }
\end{table}
 
\begin{table}[htbp]
  \centering
  \begin{tabular}{lccc}
    \hline
    \textbf{Class} & \textbf{Precision (\%)} & \textbf{Recall (\%)} & \textbf{F1-score (\%)} \\
    \hline
    0~\texttt{No Error}              & 96 & 98 & 97 \\
    1~\texttt{Logic Error}            & 95 & 93 & 94 \\
    2~\texttt{Syntax Error}           & 99 & 99 & 99 \\
    3~\texttt{Name Error}             & 90 & 98 & 93 \\
    4~\texttt{Type Error}             & 90 & 86 & 88 \\
    5~\texttt{Indentation Error}      & 100 & 99 & 99 \\
    6~\texttt{Unbound Local Error}     & 38 & 14 & 21 \\
    7~\texttt{Key Error}              & --- & --- & --- \\
    8~\texttt{Index Error}            & 100 & 50 & 67 \\
    9~\texttt{EOF Error}              & 100 & 100 & 100 \\
    10~\texttt{Recursion Error}       & 93 & 81 & 87 \\
    11~\texttt{Value Error}           & 100 & 14 & 25 \\
    12~\texttt{Tab Error}             & 71 & 100 & 83 \\
    13~\texttt{Other Errors}         & 100 & 43 & 60 \\
    \hline
    Macro avg     & 90 & 75 & 78 \\
    Weighted avg  & 95 & 95 & 95 \\
    \hline
  \end{tabular}
  \caption{\label{tab:appendix-codellama-taskC-test}
    CodeLlama Task C (Fine-grained Multi-class Classification) results on the dev set under problem-level split by \texttt{QuestionID}. Class~7 (\texttt{Key Error}) shows ``---'' because the \texttt{QuestionID}-based split yields no \texttt{Key Error} samples in the dev set. Per-class precision/recall/F1 are reported at decimal precision; Macro and Weighted averages follow the stored full-precision computation (one decimal place). Error types are indexed according to the taxonomy described in Section~\ref{sec:taxonomy-design}.
  }
\end{table}

\clearpage
\subsection{Model Performance Evaluation Metrics and Prompt Templates}

For submission $i$, let $G_i$ and $P_i$ denote the expert and predicted
label sets. Exact-set match is $\mathds{1}[G_i=P_i]$. Sample F1 is the
average of $2|G_i\cap P_i|/(|G_i|+|P_i|)$ over submissions. Micro F1 pools
all label decisions before computing precision and recall, whereas macro F1
averages the per-label F1 values over the 13 labels with gold support.

The historical experiment package also reports a \emph{contains} score. Its
legacy alignment script checks whether one normalized gold identifier occurs
in the predicted set; it does not require recovery of the complete expert
set. We therefore retain that score only for provenance and do not use it for
the current multi-error claims. Top-1 analyses use only the first parsed
prediction and treat a missing prediction as an empty set.

\paragraph{Prompt Templates.}
We list the full prompt templates used in our prompting classification experiments. Listing~\ref{lst:single-error-prompt} shows the prompt for single-error classification, and Listing~\ref{lst:multi-error-prompt} shows the prompt for multi-error classification.

\label{sec:appendix-prompt}

\begin{lstlisting}[
caption={Single-error prompt template used in the prompting classification experiments.},
xleftmargin=2em,
  xrightmargin=2em,
label={lst:single-error-prompt},
basicstyle=\ttfamily\small,
breaklines=true,
frame=single
]
# Define prompt template
prompt_template = 
You are an expert in Python code error classification, functioning as an automated assessment assistant for a large-scale programming education platform. Your task is to analyze student code submissions, which may contain errors, along with the corresponding coding questions and correct answers. You need to accurately identify whether the student code contains any errors, as well as the specific type of error in each case.

For each submission, you will receive structured information, including:
- The question description
- The expected answer
- The student's code

**Important**: Many student submissions may be completely correct with no errors.

## Analysis Steps:
1. First, see the student's code to check if the code has explicit errors (Syntax Error, Name Error, Type Error, Indentation Error, UnboundLocal Error, Key Error, Index Error, EOF Error, Value Error, Tab Error)
2. If there is no explicit errors, see the question description, the student's code, and compare with the expected answer to check the logic of code
3. If the logic is wrong, there is an logic error; Else, there is no error.

Your goal is to classify the error into a predefined taxonomy of Python error types. Please output only the label number and label name, separated by a space, e.g., "0 No error" or "3 NameError".

## Definition of Error Type Categories (Label number and names)
0: No error - The code runs successfully with no explicit or logic errors.
1: LogicError - The code has no explicit error but has a logic flaw.
2: SyntaxError - The code contains invalid Python syntax and cannot be parsed. Examples include missing colons, unmatched parentheses, etc.
3: NameError - Raised when attempting to access a variable, function, or module name that is not defined.
4: TypeError - Raised when an operation or function is applied to an object of inappropriate type or when types are incompatible.
5: IndentationError - The code has incorrect indentation. Python relies on indentation to determine code blocks.
6: UnboundLocalError - Raised when a local variable is referenced before assignment (typically within a function). This is a subclass of NameError.
7: KeyError - Raised when attempting to access a dictionary key that doesn't exist.
8: IndexError - Raised when attempting to access an index position in a sequence (like lists, tuples, or strings) that is out of range.
9: EOFError - Raised when the input() function hits an end-of-file condition without reading any data.
10: RecursionError - Raised when the maximum recursion depth is exceeded, typically caused by infinite recursion.
11: ValueError - Raised when a function receives an argument of the correct type but an inappropriate value.
12: TabError - Raised when indentation contains inconsistent use of tabs and spaces.
13: Other errors - Includes AttributeError (Raised when attempting to access an attribute or method that doesn't exist on an object), RuntimeError (A generic error raised when an error is detected that doesn't fall into any other category), SyntaxWarning (Issued when the syntax is questionable but not invalid enough to be a SyntaxError), ZeroDivisionError (Raised when division or modulo operation is performed with zero as the divisor), MemoryError (Raised when an operation runs out of memory), ModuleNotFoundError (Raised when a module could not be found. This is a subclass of ImportError), etc.
\end{lstlisting}

\begin{lstlisting}[
caption={Multi-error prompt template used in the prompting classification experiments.},
xleftmargin=2em,
  xrightmargin=2em,
label={lst:multi-error-prompt},
basicstyle=\ttfamily\small,
breaklines=true,
frame=single
]
prompt template = 
You are an expert in Python code error classification, functioning as an automated assessment assistant for a large-scale programming education platform. Your task is to analyze student code submissions, which may contain errors, along with the corresponding coding questions and correct answers. You need to accurately identify whether the student code contains any errors, as well as the specific type of error in each case.

For each submission, you will receive structured information, including:
- The question description
- The expected answer
- The student's code

**Important**: Many student code submissions may be completely correct with no errors.

---
#Analysis Process (internal reasoning, not directly output):
1. Examine the student's code for explicit errors (Syntax Error, Name Error, Type Error, Indentation Error, Unbound Local Error, Key Error, Index Error, EOF Error, Recursion Error, Value Error, Tab Error, or Other Errors).
2. Logic Error check: Logic Error may co-occur with explicit errors. Assume that all explicit errors identified in Step 1 are minimally repaired in a way consistent with the apparent intent. Add Logic Error only if the repaired code would still violate the question. This judgment must be supported by a concrete valid input, edge case, output requirement, or required behavior. If the repaired code would be correct, do not add Logic Error.
3. If neither an explicit error nor a Logic Error is found, classify as No Error. No Error must not co-occur with another label.
Use this reasoning process internally to justify the final prediction.

---
#Definition of Error Type Categories (Label numbers and names)
0: No Error: The code runs successfully with no explicit or Logic Error.
1: Logic Error: After minimally repairing identified explicit errors, the code still violates a concrete requirement of the question.
2: Syntax Error: The code contains invalid Python syntax and cannot be parsed.
3: Name Error: Raised when attempting to access a variable, function, or module name that is not defined.
4: Type Error: Raised when an operation or function is applied to an object of inappropriate type or when types are incompatible.
5: Indentation Error: Incorrect indentation. Python relies on indentation to determine code blocks.
6: Unbound Local Error: A local variable is referenced before assignment (typically within a function). Subclass of Name Error.
7: Key Error: Raised when accessing a dictionary key that doesn't exist.
8: Index Error: Raised when accessing an index position in a sequence that is out of range.
9: EOF Error: Raised when the input() function hits an end-of-file condition without reading data.
10: Recursion Error: Raised when the maximum recursion depth is exceeded (infinite recursion).
11: Value Error: Raised when an argument has the correct type but an inappropriate value.
12: Tab Error: Raised when indentation inconsistently mixes tabs and spaces.
13: Other Errors: Includes AttributeError, RuntimeError, SyntaxWarning, ZeroDivisionError, MemoryError, ModuleNotFoundError, etc.

---
## Output Requirements
Please provide output in the following format (consistent and unified):
Reasoning: brief explanation of detected errors
Predicted Label: ErrorName (#n), ErrorName (#m), ...

---
## Example
### Input:
Question description: "Write a function that returns the square of a number."

Expected Answer:
def square(x):
    return x * x

Student Code:
def square(x)
    return x + y

Output:
Reasoning: Firstly, we examine the code for explicit errors in order. The function header is missing a colon (Syntax Error). The variable y is not defined (Name Error). While there are explicit errors, we continue to examine the logic. Even if y were defined, the logic is incorrect since it should return x * x instead of x + y (Logic Error).
Predicted Label: Syntax Error (#2), Name Error (#3), Logic Error (#1)

Test Example (for model to fill)
Reasoning:
Predicted Label:
\end{lstlisting}

\subsection{Single-Error Prompting Classification: Earlier Prompted Models}
\label{sec:appendix-single-prompting}

\begin{table}[htbp]
  \centering
  \resizebox{0.95\textwidth}{!}{%
  \begin{tabular}{lccccc}
    \hline
    \textbf{Model}
    & \texttt{\textbf{No Error (\%)}}
    & \texttt{\textbf{Indentation Error (\%)}}
    & \texttt{\textbf{Logic Error (\%)}}
    & \texttt{\textbf{Name Error (\%)}}
    & \texttt{\textbf{Type Error (\%)}} \\
    \hline
    GPT-3.5 & 34.7 & 0.0  & 92.3 & 3.3  & 0.0 \\
    GPT-4o  & 85.6 & 36.8 & 76.2 & 36.2 & 18.3 \\
    Gemini   & 83.2 & 84.1 & 89.6 & 96.3 & 91.0 \\
    DeepSeek-V3 & 81.9 & 46.8 & 85.2 & 73.5 & 26.2 \\
    \hline
  \end{tabular}
  }
  \caption{Single-Error Prompting Classification: Per-class accuracy (in \%) for selected error types across the earlier prompted models. ``Gemini'' denotes Gemini 2.5 Pro.}
  \label{tab:per_class_accuracy}
\end{table}

Table~\ref{tab:prompting-single} summarizes the overall performance of four representative models.

\begin{table}[htbp]
  \centering
  \begin{tabular}{lcc}
    \hline
    \textbf{Model} & \textbf{Runtime} & \textbf{Accuracy (\%)} \\
    \hline
    GPT-3.5        & 45 mins     & 40.3 \\
    GPT-4o         & 45 mins     & 71.5 \\
    Gemini     & 24 hours    & 85.9 \\
    DeepSeek-V3       & 5--6 hours  & 73.6 \\
    \hline
  \end{tabular}
  \caption{\label{tab:prompting-single}
    Single-Error Prompting Classification of Runtime and Accuracy. Overall accuracy and approximate inference runtime for four LLMs evaluated on the full test set (4,865 samples).
  }
\end{table}

Inference time in the earlier set of prompted models ranges from 45 minutes to 24 hours (Table~\ref{tab:prompting-single}). Gemini 2.5 Pro obtains the highest accuracy, 85.9\%, and takes 24 hours. GPT-4o reaches 71.5\% in 45 minutes, while DeepSeek-V3 reaches 73.6\% in 5--6 hours. GPT-3.5 also takes 45 minutes but reaches only 40.3\%.

In this earlier set of prompted models, all four models tend to over-predict
\texttt{Logic Error}. This pattern is much smaller for the recent set of prompted models in
the main evaluation. Per-class accuracy for the earlier models appears in
Table~\ref{tab:per_class_accuracy}, and the comparison between the two sets of prompted models appears
in Table~\ref{tab:model-performance}.

\begin{figure}
    \centering
    \includegraphics[width=1\linewidth]{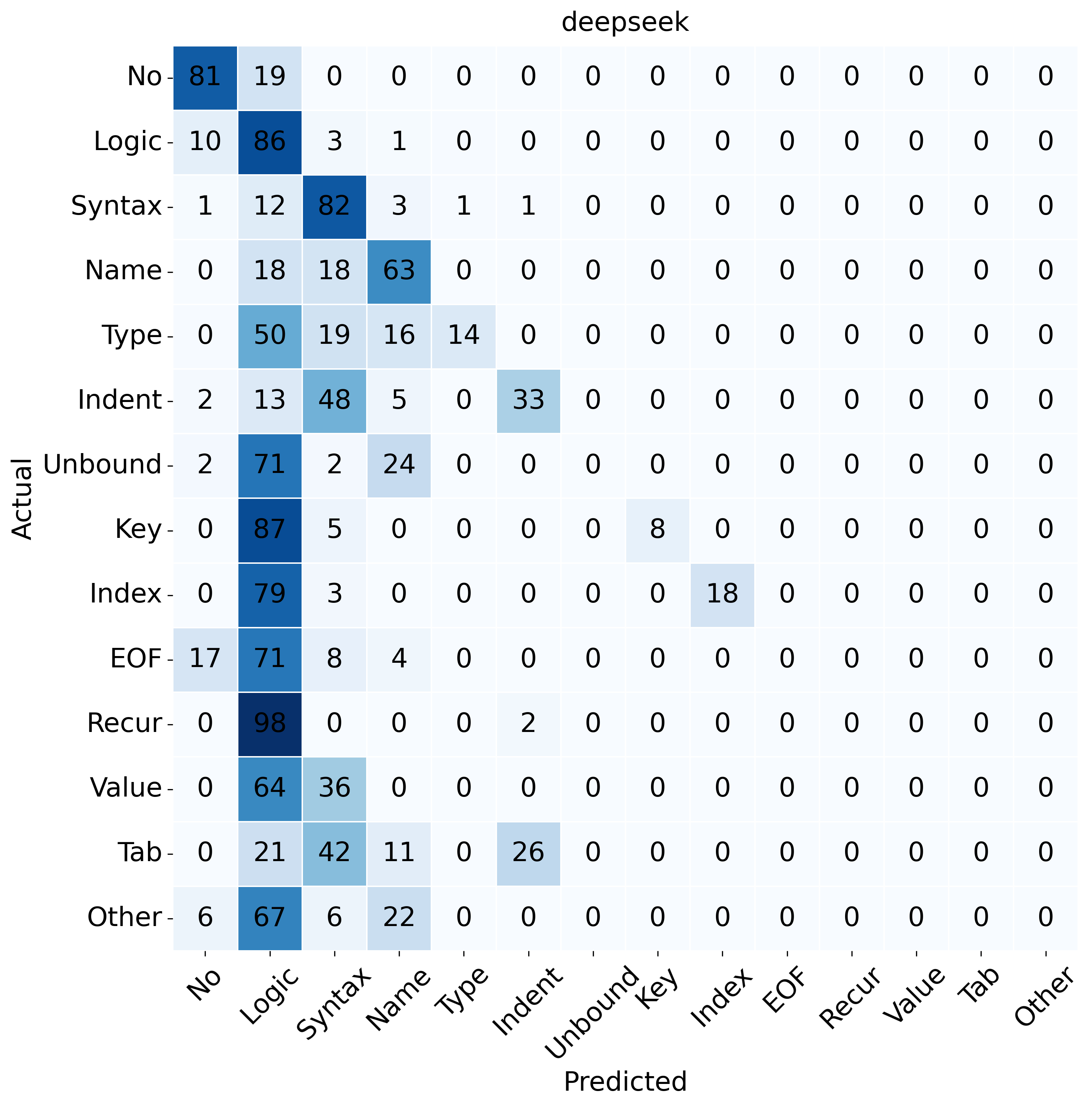}
    \caption{DeepSeek-V3 Row-Normalized Confusion Matrix of Single-Error Prompting Classification.}
    \label{fig:confusion_DeepSeek-V3}
\end{figure}
\begin{figure}
    \centering
    \includegraphics[width=1\linewidth]{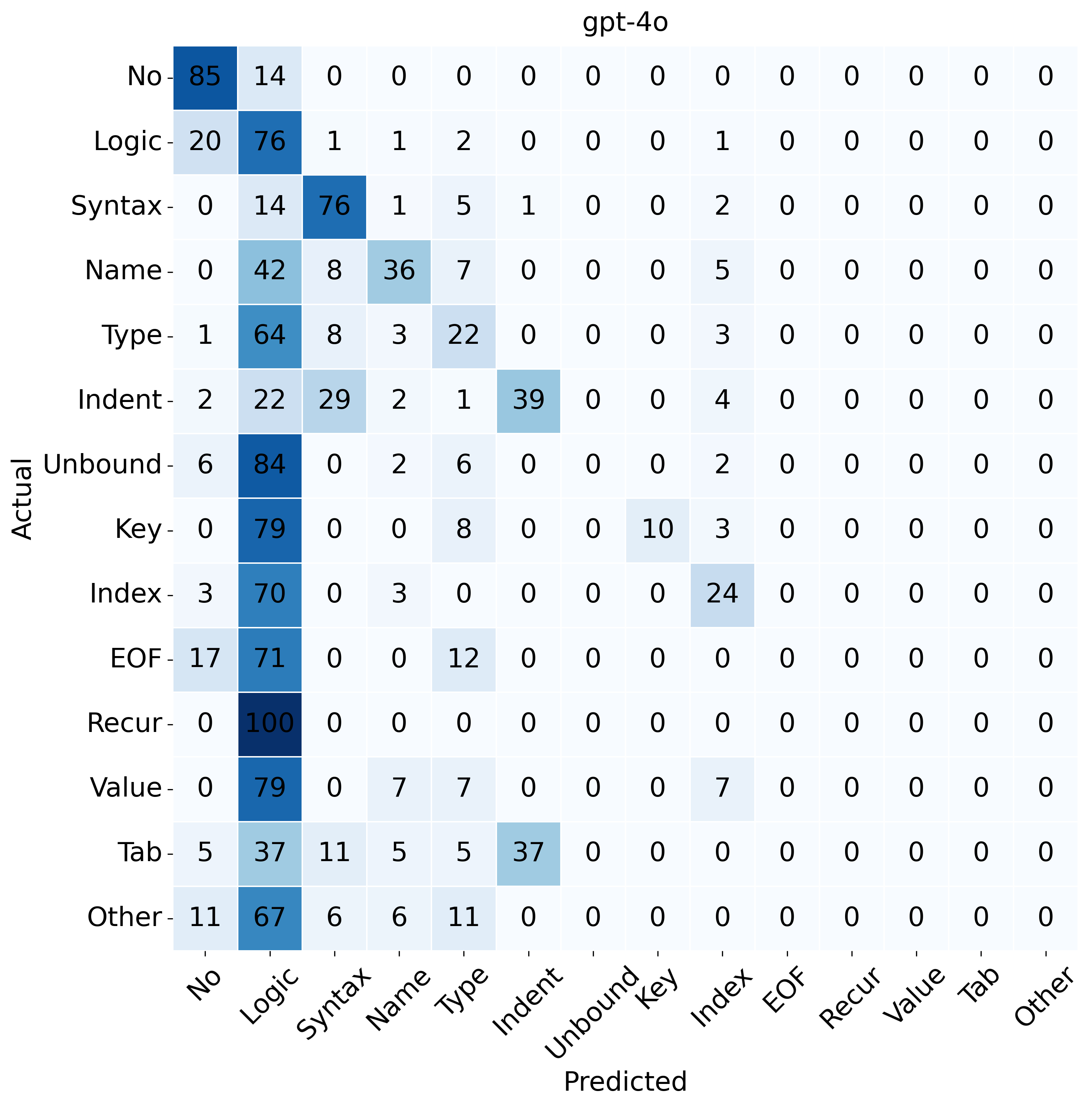}
    \caption{GPT-4o Row-Normalized Confusion Matrix of Single-Error Prompting Classification.}
    \label{fig:confusion_gpt4o}
\end{figure}

In the per-class results, Gemini 2.5 Pro performs best on
\texttt{Indentation Error}, \texttt{Logic Error}, \texttt{Name Error}, and
\texttt{Type Error}. DeepSeek-V3 is strongest on \texttt{Logic Error} and
\texttt{Name Error} among its own class results, whereas GPT-4o performs
poorly on those two classes.

\subsubsection{Qualitative Analysis of SE$\rightarrow$LE Misclassification}
\label{app:se-le-qualitative}

To complement the quantitative results reported in Section~\ref{sec:single-error-prompting-classification},
we inspect representative cases in which
\texttt{Syntax Error} instances are predicted as \texttt{Logic Error}
(SE$\rightarrow$LE).

We find that such cases often involve code with syntactic errors whose
surrounding structure still suggests what the student was trying to do.
In these cases, models tend to pick up on the intended logic rather than
flag the syntactic problem. This is consistent with the intent-first bias
discussed in the main text.

Single-error gold labels follow the IDE convention of
recording the first execution error. When a submission contains multiple
co-occurring issues, the gold label is therefore fixed to the first
execution error, even if a model could also reasonably identify another
error in the code. The diagnostic multi-error subset
(Section~\ref{sec:multi-error-diagnostic}) is designed precisely to
examine model behavior under such conditions.

A possible contributing factor is the co-occurrence of explicit and logic errors, particularly in cases where logic-related anomalies appear earlier in the code.
Although the model is explicitly instructed to consider \texttt{Logic Error} only in the absence of other error types, these examples suggest that the model does not always strictly adhere to this constraint. This behavior may stem from imperfect adherence to prompt instructions, incomplete recognition of co-occurring errors, or an implicit bias toward logic errors. Distinguishing among these mechanisms requires further investigation.

Two representative examples are shown below in Section~\ref{sssec:rep_examples}. In both cases, the code
contains a syntactic error that would prevent execution, but the
surrounding code makes the intended computation easy to read off,
and the model returns a \texttt{Logic Error} label instead.

\subsubsection{Reasoning and Overprediction Case Studies}

To provide concrete illustrations of the observed biases and uncertainty, we conduct qualitative case studies on reasoning failures and Logic Error Overpredictions.

\subsubsection{Representative Examples}
\label{sssec:rep_examples}

\begin{lstlisting}[language=Python,
caption={Example 1. A submission containing invalid assignment expressions (\texttt{Syntax Error}) misclassified as \texttt{Logic Error}.},
xleftmargin=2em,
  xrightmargin=2em,
label={lst:296600}
]
input_str = input()
position = {}
(*@\mbox{\textcolor{red}{\bfseries\ttfamily str(res) = \textquotesingle\textquotesingle}}@*)  # invalid target
(*@\mbox{\textcolor{red}{\bfseries\ttfamily int(pos) = len(input\_str)}}@*)  # invalid target

def get_first_duplicate(input_str):
    str(input_str)
    for i in input_str:
        if i not in position:
            position[i] = input_str.index(i)
        else:
            if input_str.index(i) < pos:
                res = i
                pos = input_str.index(i)
    return res

print(get_first_duplicate(input_str))
\end{lstlisting}

\begin{lstlisting}[language=Python,
caption={Example 2. A submission with malformed expressions (\texttt{Syntax Error}) in a near-correct computation.},
xleftmargin=2em,
  xrightmargin=2em,
label={lst:49484}
]
age = int(input())
weight = int(input())
heart_rate = int(input())
time = int(int(input()))

calories_burned = (age + weight + heart_rate + time - 75.4991) (*@\color{red}\textbf{\textbackslash}@*)
    * time / (8.368 / age * 0.2757) (*@\color{red}\textbf{\textbackslash}@*)
    / (8.368 / weight * 0.03295) (*@\color{red}\textbf{\textbackslash}@*)
    / (8.368 / heart_rate * 1.0781)

print(calories_burned)
\end{lstlisting}

\begin{lstlisting}[language=Python,
caption={Example 3. A submission with logical redundancy where a nested \texttt{elif} condition replicates its parent \texttt{if} statement.},
xleftmargin=2em,
  xrightmargin=2em,
label={lst:spider_case}
]
num_spiders = int(input())
num_cats = int(input())
if num_spiders > 10:
    if num_cats > num_spiders:
        print('cat will eat spiders')
    (*@\color{red}\textbf{elif num\_spiders > 10:}@*) # Logical Redundancy (Dead Code)
        if num_cats <= num_spiders:
            print('cat will not eat all spiders')
else:
    print('few spiders')
\end{lstlisting}

\subsubsection{Behavioral Alignment Between Model Explanations and Actual Execution Behavior}
\label{appendix:behavioral-alignment}

\paragraph{Scope of this analysis}
The primary goal of our prompting experiments is to evaluate classification performance and
identify systematic model biases, not to analyse internal execution-reasoning mechanisms.
Nevertheless, our prompting setup instructs models to output a brief \textit{Reasoning}
string alongside each predicted label (see Appendix~\ref{sec:appendix-prompt}).
We use these model-generated explanations as a behavioural signal: by comparing what a
model \emph{says} about a submission against what the Python interpreter
\emph{actually does} when it processes that code, we can assess whether the model's
decision is aligned with real execution behaviour.
Concretely, Python errors follow a strict execution order:
\emph{parse-time failures} (e.g.\ \texttt{Syntax Error}) occur before any code runs;
\emph{runtime exceptions} (e.g.\ \texttt{NameError}, \texttt{TypeError}) occur during
execution; and \emph{logic failures} occur only after the code runs to completion but
produces wrong output.
A prediction is \textbf{execution-aligned} if the model's explanation is consistent with
the stage at which the error actually occurs.

\paragraph{Observed patterns}
Inspecting the model explanations for misclassified samples, we identify two
systematic misalignment patterns.

\textbf{Pattern~1: Intent-first reasoning (SE to LE).}
When a submission contains a parse-level error, the interpreter halts immediately and
never evaluates any logic.
Yet models frequently skip the parse stage and reason directly about algorithmic
correctness, producing a \texttt{Logic Error} prediction for code that cannot run at all.

\begin{lstlisting}[language=Python,
  caption={Leap-year check containing a full-width colon \texttt{\cjk{：}} (line~2).
           The parser fails at this character; no logic is ever evaluated.},
  label={lst:leapyear}]
def is_leap_year(year):
    if year % 400 == 0(*@\cjk{：}@*)      # full-width colon -- parse fails here
        return "leap year"
    elif year % 100 == 0:
        return "not a leap year"
    elif year % 4 == 0:
        return "leap year"
\end{lstlisting}

\noindent
\textbf{True label:} \texttt{Syntax Error} (parse-time failure on the full-width colon).\\
\textbf{Model prediction:} \texttt{Logic Error}.\\
\textbf{Model explanation:} \textit{``the leap year conditions are not correctly
structured \ldots\ the order of the conditions is incorrect.''}

The explanation evaluates the correctness of the conditional logic , a question that
is only meaningful if the code can be parsed and executed.
Because the parser fails on line~2, the logical structure is never reached.
The model's reasoning is grounded in the code's \emph{intended behaviour}, not in
what the interpreter encounters.

A second case shows the same pattern more starkly:

\begin{lstlisting}[language=Python,
  caption={Print statement with no quotation marks around the argument.
           The interpreter raises a \texttt{Syntax Error}/\texttt{NameError}
           immediately; the statement cannot be parsed as written.},
  label={lst:print}]
print(I am Python!)
\end{lstlisting}

\noindent
\textbf{True label:} \texttt{Syntax Error}.\\
\textbf{Model prediction:} \texttt{Logic Error}.\\
\textbf{Model explanation:} \textit{``the issue lies in the logic of the code \ldots\
the incorrect logic of not enclosing the string within quotation marks.''}

The model reframes a structural parse failure as a reasoning mistake by the student,
describing the problem in terms of intent rather than interpreter state.

\textbf{Pattern~2: Surface-feature substitution (LE to SE).}
In the reverse direction, models sometimes predict \texttt{Syntax Error} for code that
is syntactically valid and executes without error.
The explanations cite visible surface features , unconventional spacing, unfamiliar
constructs, or naming inconsistencies , as evidence of a syntax violation.

\begin{lstlisting}[language=Python,
  caption={Discount function that runs without error but returns strings instead of
           integers, giving wrong output (\texttt{Logic Error}).},
  label={lst:discount}]
def get_discount(param):
    if param < 140:
        return '0'       # executes fine; wrong type
    else:
        return'100'      # missing space, but valid Python syntax
\end{lstlisting}

\noindent
\textbf{True label:} \texttt{Logic Error} (returns \texttt{`0'}/\texttt{`100'} instead
of \texttt{0}/\texttt{100}; code runs without exception).\\
\textbf{Model prediction:} \texttt{Syntax Error}.\\
\textbf{Model explanation:} \textit{``\texttt{return`100'} lacks a space, violating
PEP~8 \ldots\ looks broken to automated tools; the parameter name \texttt{param} does
not match the expected signature \texttt{height} , a static mismatch.''}

Neither observation prevents execution: Python does not require whitespace between
\texttt{return} and its argument, and parameter naming does not affect runtime
behaviour.
The model's prediction is driven by static features that \emph{resemble} syntax
problems rather than by whether the code actually executes.

\textbf{Pattern~3: Naming a fault behind the first outcome (recent prompted models).}
The single-error prompt asks for the label only
(Listing~\ref{lst:single-error-prompt}), so no explanation text exists for the
recent prompted models (GPT-5.5, Gemini 3.1 Pro Preview, Claude Opus 5, and
DeepSeek-V4-Pro). For these models we compare the predicted label with the
interpreter's first outcome and use agreement of all four models on the same wrong
label as the behavioural signal. Two test-split submissions show a pattern that the
explanations of the earlier models did not expose: the predicted label names a fault
that is present in the program but is not the first outcome the interpreter produces.

\begin{lstlisting}[language=Python,
  caption={Submission that reads standard input twice although the task supplies one
           value. The second \texttt{input()} raises \texttt{EOFError} before the
           \texttt{str + int} addition on the same line can raise \texttt{TypeError}.},
  label={lst:karma}]
karma = int(input())
karma = int(input()+1)   # EOF here; the str + 1 TypeError is never reached
print(karma)
\end{lstlisting}

\noindent
\textbf{True label:} \texttt{EOFError} (the judge supplies a single input line, so the
second \texttt{input()} fails on line~2).\\
\textbf{Model prediction:} \texttt{TypeError}, returned by all four recent prompted
models.

The predicted error is real: once a second input line is supplied,
\texttt{input()+1} raises \texttt{TypeError}. Under the first-outcome protocol the
label is nevertheless \texttt{EOFError}, because the interpreter stops at the earlier
exception. All four models name the fault that would surface after the first repair,
which is the kind of label the repair-conditional expert sets of
Section~\ref{sec:multi-error-diagnostic} are designed to record.

\begin{lstlisting}[language=Python,
  caption={Off-by-one index for a task that asks for the fifth character
           (\texttt{param[4]}). On the example input \texttt{CircleCat} the program
           prints \texttt{e} instead of \texttt{l}; on a five-character hidden test it
           raises \texttt{IndexError}.},
  label={lst:index}]
param = input()
print(param[5])          # the fifth character is param[4]
\end{lstlisting}

\noindent
\textbf{True label:} \texttt{IndexError} (the third hidden test is the five-character
string \texttt{Happy}; a runtime exception on any test takes precedence over wrong
output on the others).\\
\textbf{Model prediction:} \texttt{Logic Error}, returned by all four recent prompted
models. The test split contains six submissions to this task that index the input at a
fixed position; each of the four models labels five of them \texttt{Logic Error}.

Here the models' reading is correct on the information they receive. They see the
problem statement, the reference answer, and the submission
(Section~\ref{sec:single-error-prompting}), not the hidden tests, and on the visible
example the program runs and prints the wrong character. The gold label is decided by
an input the model never sees. Label-level evidence cannot distinguish a model that
considered short inputs and dismissed them from one that never considered them; the
first-outcome score records both as the same error.

\paragraph{General trend}
Across the first two patterns, model explanations consistently focus on
\emph{apparent structure and intent} rather than on the execution stage at which the
error occurs.
In SE$\rightarrow$LE cases, models reason about logic that is never reached.
In LE$\rightarrow$SE cases, models flag surface features that do not prevent execution.
This behavioural misalignment is consistent with the quantitative bias reported in
Section~\ref{sec:logic-error-bias}: models over-predict \texttt{Logic Error} because they tend
to reason about code intent first, and under-predict fine-grained explicit errors
because surface-level syntactic cues are ambiguous.
The third pattern, observed at the label level for the recent prompted models, is
different in kind: the predicted label names a fault that is present in the program
but is not the first outcome, either because an earlier exception pre-empts it
(Listing~\ref{lst:karma}) or because the triggering input is a hidden test
(Listing~\ref{lst:index}).
Both predictions count as errors under the first-outcome protocol. Whether either
would be supported along a repair path can only be settled by the kind of expert audit
used in Section~\ref{sec:multi-error-diagnostic}; neither submission is in the
97-item subset.
We emphasise that the analysis above is based solely on model-generated explanations
(Patterns~1 and~2) and on predicted labels (Pattern~3), not on controlled execution
traces or step-by-step interpreter simulation.
A more rigorous execution-grounded evaluation, for example, prompting models to
explicitly simulate interpreter state at each line before producing a label, similar
to the approach of \citet{wang2025semantics}, would provide stronger evidence about
whether models genuinely track execution behaviour.
We leave this as a direction for future work.

\subsection{Legacy Multi-Error Results: Earlier Prompted Models}

The following table is retained to document the earlier set of prompted models and
its evaluation script. As explained in Appendix~\ref{sec:appendix-prompt},
the legacy \emph{contains} score is not a complete-set metric and is not used
for the current RQ3 claims.

\begin{table}[htbp]
  \centering
  \begin{tabular}{llccc}
    \hline
    \textbf{Dataset} & \textbf{Model} & \textbf{Precision (\%)} & \textbf{Recall (\%)} & \textbf{F1 (\%)} \\
    \hline
    \multicolumn{5}{l}{\textbf{Total (215 + 97 samples)}} \\
    \hline
     & DeepSeek-V3  & 85.7 & 49.0 & 60.1 \\
     & Gemini    & 85.7 & 62.0 & 70.2 \\
     & GPT-3.5   & 71.4 & 30.1 & 39.9 \\
     & GPT-4o    & 64.3 & 39.9 & 46.6 \\
    \hline
    \multicolumn{5}{l}{\textbf{Multi-Error Subset (97 samples)}} \\
    \hline
     & DeepSeek-V3  & 84.6 & 47.2 & 58.2 \\
     & Gemini    & 84.6 & 59.9 & 68.4 \\
     & GPT-3.5   & 69.2 & 30.1 & 39.4 \\
     & GPT-4o    & 61.5 & 36.6 & 43.2 \\
    \hline
    \multicolumn{5}{l}{\textbf{\texttt{No Error} Subset(134 samples)}} \\
    \hline
     & DeepSeek-V3  & 100.0 & 75.0 & 85.7 \\
     & Gemini    & 100.0 & 88.2 & 93.8 \\
     & GPT-3.5   & 100.0 & 31.6 & 48.0 \\
     & GPT-4o    & 100.0 & 86.0 & 92.5 \\
    \hline
    \multicolumn{5}{l}{\textbf{\texttt{Logic Error} Subset(81 samples)}} \\
    \hline
     & DeepSeek-V3  & 100.0 & 90.2 & 94.9 \\
     & Gemini    & 100.0 & 93.9 & 96.9 \\
     & GPT-3.5   & 100.0 & 78.0 & 87.7 \\
     & GPT-4o    & 100.0 & 85.4 & 92.1 \\
    \hline
  \end{tabular}
  \caption{\label{tab:multi-error-summary}
    Legacy multi-error results for the earlier set of prompted models. The table
    reproduces values emitted by the historical \emph{contains}-mode script;
    these values should not be interpreted as complete-set recovery. The top
    block combines 215 single-label cases and the 97-case diagnostic subset;
    the remaining blocks show the historical partitions. All models were
    evaluated with prompting and without finetuning.
  }
\end{table}

\subsection{Entropy--F1 Relationship Analysis}
\label{sec:appendix-entropy-f1}

To examine how model uncertainty relates to human judgments in multi-error settings, we align each of the 97 expert annotated multi-error sample with its prediction entropy and top-$k$ model output. Therefore, each sample contains both the human-identified error set and the ranked error-type predictions of the model. 
We compute entropy from the model's distribution over error types and group samples into entropy buckets. Within each bucket, we compare the model's top-$k$ predictions with the expert labels using macro F1. Agreement decreases as entropy rises. Because the subset was partly selected for high entropy and each bucket is small, this analysis describes the selected cases; it does not estimate a general relationship between entropy and expert agreement.

\begin{figure}
    \centering
    \includegraphics[width=1\linewidth]{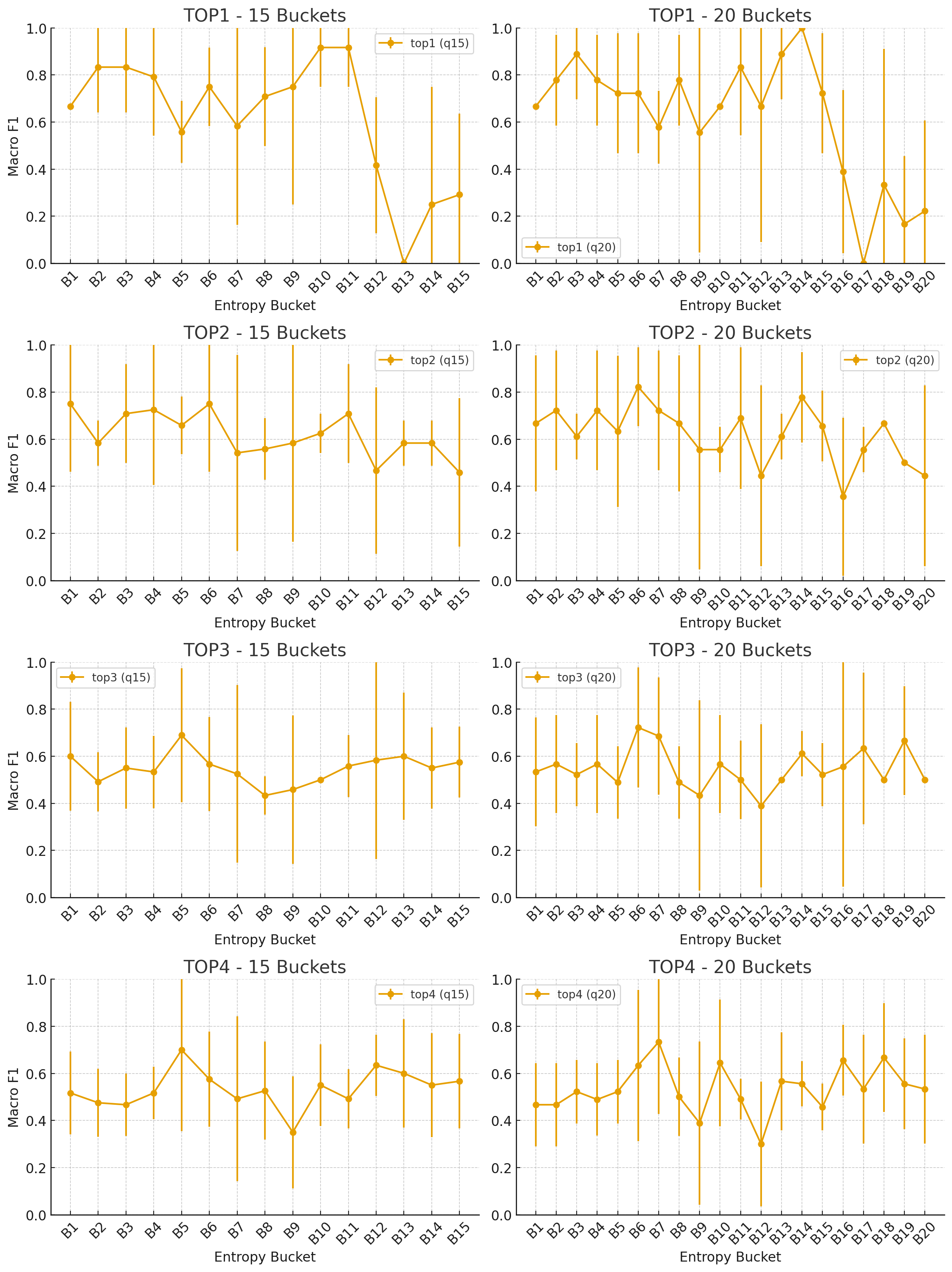}
    \caption{Entropy--F1 trend in the multi-error diagnostic analysis. Macro
    F1 of top-$k$ model predictions ($k=1$--$4$) across entropy buckets
    (15- and 20-bucket settings) on the 97-case diagnostic subset. Results
    use single-error finetuning (CodeBERT Task C); entropy is computed from
    the predicted class distribution.}
    \label{fig:entropy_f1_trend}
\end{figure}
Figure~\ref{fig:entropy_f1_trend} illustrates the Macro-F1 scores across entropy buckets 
for \textit{Top1–Top4} predictions (15- and 20-bucket settings). 
A clear trend is observed: as entropy increases, Macro-F1 decreases, 
confirming our hypothesis that higher uncertainty correlates with lower prediction reliability. 
This downward trend is most pronounced for \textit{Top1} predictions, 
less distinct for \textit{Top2}, and nearly disappears for \textit{Top3–Top4}. 
The results indicate that including additional top-$k$ candidates 
smooths the performance degradation across entropy, 
suggesting that while the model’s most confident predictions degrade sharply with uncertainty, 
its broader candidate set remains more stable.

The standard deviation of Macro-F1 within each bucket also increases with entropy, 
implying greater variability in model performance on uncertain samples.

\subsection{Use of AI Assistants}
\label{sec:ai-assistants}

AI assistants were used for language editing, document organization, and coding assistance during metric verification. The authors determined the research questions and methods, verified all reported results against saved experiment outputs, and take responsibility for the manuscript. The evaluated model outputs are part of the experiments described in the paper and are not used as ground-truth annotations.

\end{document}